\documentclass[preprint]{aastex702}
\newcommand{\update}[1]{\textcolor{black}{#1}}
\usepackage{CJKutf8}
\usepackage{graphicx}
\usepackage{amsmath}

\begin{document}

\title{NSF-DOE Vera C. Rubin Observatory's First View of the Neptune Trojan L5 (Trailing) Cloud: Two New Long-Term Stable Trojans}

\author[orcid=0000-0003-4365-1455,gname=Megan,sname=Schwamb]{Megan E. Schwamb}
\affiliation{Astrophysics Research Centre, School of Mathematics and Physics, Queen's University Belfast, Belfast BT7 1NN, UK}
\email[show]{m.schwamb@qub.ac.uk}  
\correspondingauthor{Megan E. Schwamb}

\author[orcid=0000-0003-0743-9422, gname=Pedro, sname=Bernardinelli]{Pedro H. Bernardinelli} 
\affiliation{Departamento de Astronomia, Instituto de Astronomia, Geofísica e Ciências Atmosféricas, Universidade de S\~ao Paulo, 05508-090, São Paulo, SP, Brazil} 
\email{pedro.bernardinelli@usp.br}  

\author[orcid=0000-0001-8736-236X, gname=Kathryn, sname=Volk]{Kathryn Volk}
\affiliation{Planetary Science Institute, 1700 East Fort Lowell, Suite 106, Tucson, AZ 85719, USA}
\email{kvolk@psi.edu} 

\author[orcid=0000-0001-9505-1131, gname=Joseph, sname=Murtagh]{Joseph Murtagh} 
\affiliation{DiRAC Institute and the Department of Astronomy, University of Washington, 3910 15th Ave NE, Seattle, WA 98195, USA} 
\email{murtagh@uw.edu}

\author[orcid=0000-0001-7737-6784, gname=Hsing~Wen, sname=Lin]{Hsing~Wen~Lin (\begin{CJK*}{UTF8}{bkai}林省文\end{CJK*})}
\affiliation{Department of Physics, University of Michigan, Ann Arbor, MI 48109, USA}
\affiliation{Michigan Institute for Data and AI in Society, University of Michigan, Ann Arbor, MI 48109, USA}
\affiliation{Department of Physics, Case Western Reserve University, Cleveland, OH 44106, USA}
\email{hsing-wen.lin@case.edu}

\author[orcid=0000-0002-9939-9976, gname=Andrew, sname=Rivkin]{Andrew S. Rivkin}
\affiliation{Johns Hopkins University Applied Physics Laboratory, Laurel, Maryland, USA}
\email{Andy.Rivkin@jhuapl.edu}  

\author[orcid=0000-0001-7335-1715, gname=Colin, sname=Chandler]{Colin Orion Chandler}
\affiliation{DiRAC Institute and the Department of Astronomy, University of Washington, 3910 15th Ave NE, Seattle, WA 98195, USA}
\email{coc123@uw.edu}  

\author[orcid=0000-0001-5916-0031, gname=R. Lynne, sname=Jones]{R. Lynne Jones}
\affiliation{NSF-DOE Vera C. Rubin Observatory, 950 N. Cherry Ave., Tucson, AZ 85719, USA}
\affiliation{Aston Carter, Suite 150, 4321 Still Creek Drive, Burnaby, BC V5C6S, Canada}
\email{ljones.uw@gmail.com}  

\author[orcid=0000-0001-8821-5927, gname=Susan, sname=Benecchi]{Susan D. Benecchi}
\affiliation{Planetary Science Institute, 1700 East Fort Lowell, Suite 106, Tucson, AZ 85719, USA}
\email{susank@psi.edu}

\author[0000-0003-1996-9252, gname=Mario, sname=Juri\'c]{Mario Juri\'c}
\affiliation{DiRAC Institute and the Department of Astronomy, University of Washington, 3910 15th Ave NE, Seattle, WA 98195, USA}
\email{mjuric@uw.edu}  

\author[orcid=0009-0005-5452-0671, gname=Jacob, sname=Kurlander]{Jacob A. Kurlander} 
\affiliation{DiRAC Institute and the Department of Astronomy, University of Washington, 3910 15th Ave NE, Seattle, WA 98195, USA} 
\email{jkurla@uw.edu}  

\author[orcid=0000-0002-4439-1539, gname=Sarah, sname=Greenstreet]{Sarah Greenstreet}
\affiliation{NSF National Optical-Infrared Astronomy Research Laboratory, 950 North Cherry Avenue, Tucson, AZ 85719, USA}
\affiliation{DiRAC Institute and the Department of Astronomy, University of Washington, 3910 15th Ave NE, Seattle, WA 98195, USA}
\email{sarah.greenstreet@noirlab.edu}

\author[orcid=0009-0001-2321-3784,gname=Adam,sname=Wilson]{Adam Wilson}
\affiliation{Astrophysics Research Centre, School of Mathematics and Physics, Queen's University Belfast, Belfast BT7 1NN, UK}
\email{awilson110@qub.ac.uk}  

\author[0009-0003-3171-3118, gname=Maxine, sname=West]{Maxine West}
\affiliation{DiRAC Institute and the Department of Astronomy, University of Washington, 3910 15th Ave NE, Seattle, WA 98195, USA}
\email{maxwest@uw.edu}  

\author[orcid=0009-0007-8602-2954, gname=Ryan, sname=Lyttle]{Ryan R. Lyttle}
\affiliation{Astrophysics Research Centre, School of Mathematics and Physics, Queen's University Belfast, Belfast BT7 1NN, UK}
\email{rlyttle09@qub.ac.uk}

\begin{abstract}


The NSF-DOE Vera C. Rubin Observatory's combination of survey speed, sky coverage, and depth provides a unique opportunity to study the distant Trans-Neptunian region, including the Neptune Trojans. The Neptune Trojans are small bodies on orbits with Neptune's semimajor axis that are locked in the 1:1 mean motion resonance librating about the ice giant's stable Lagrange points,
L4 (leading) and L5 (trailing), located 60$^{\circ}$ ahead of or behind the planet. For more than 15 years, the on-sky location of the L5 Neptune Trojan cloud has been coincident with the Galactic plane and Galactic center; the high stellar density has significantly hindered the discovery of L5 Neptune Trojans. In 2025, the L5 cloud has now moved sufficiently out of the galactic plane to overlap with the sky covered by Rubin Observatory's Science Validation (SV) survey that was executed during the Observatory's commissioning phase.   We report on two new confirmed long-term stable (primordial) members of the Neptune L5 Trojan cloud (2025 MH348 and 2025 NN80), doubling the known sample, and one candidate (2025 MD138) that have been found in a preliminary search of the SV survey observations. We present our study to characterize the physical and dynamical properties of these objects and compare the L5 sample to the well-studied L4 cloud. \update{This work also presents the first characterization of objects discovered by Rubin Observatory's moving object discovery pipelines using observations taken in an LSST-like cadence and as such, provides a blueprint for LSST-era analyses.}

\end{abstract}



\section{Introduction} 


The Neptune Trojans are small icy bodies that orbit on the same orbit as Neptune, but they are located in swarms or clouds ahead of or behind the planet. These objects librate about the ice giant's L4 (leading) and L5 (trailing) Lagrange  points. The Neptune Trojans can be on orbits that are stable on the order of several gigayears \citep{2002Icar..160..271N, 2003A&A...410..725M, Zhou2009, 2011MNRAS.412..537L}. These long-term stable Neptune Trojans are primordial and not recent interlopers from the present-day configuration of the solar system \citep{2013Sci...341..994A, 2021PSJ.....2..212A}. The surface properties, composition, albedos, sizes, rotation rates, and orbits of these primordial Neptune Trojans provide key insights into the conditions within the early solar system and probe the conditions within the planetesimal disk that Neptune migrated through \citep{1994Icar..112..465M, 2005Natur.435..462M, 2009MNRAS.398.1715L, 2009AJ....137.5003N, 2010MNRAS.405.1375L, 2015Icar..247..112P, 2016A&A...592A.146G, 2016MNRAS.458.4277C}. 

\update{For every known Neptune Trojan, there are approximately 450 Jupiter Trojans (small bodies orbiting in the 1:1 mean motion resonance, MMR, with Jupiter) currently listed in the Minor Planet Center (MPC)\footnote{\url{https://minorplanetcenter.net}}. Although the Jupiter Trojans have been much better sampled in past surveys with nearly 16,000 discoveries\footnote{\url{https://minorplanetcenter.net/iau/lists/JupiterTrojans.html}}, the Neptune Trojans significantly outnumber Jupiter Trojans in the solar system. Past searches and surveys estimate that the Neptune Trojan population is on the order of 4-10 times larger than their counterpart at Jupiter \citep{Sheppard2006,2010ApJ...723L.233S,2012AJ....144...23G, 2016AJ....152..111A, 2019Icar..321..426L,Lin2021}. Optical surface colors of Neptune Trojans measured from broadband photometry overlap the Jupiter Trojans \citep{Sheppard2006}, but the Neptune Trojan population contains much redder surfaces not present at Jupiter's orbit \citep{2019Icar..321..426L,bolinKeckGeminiPalomar2023,markwardt2023, bernardinelli2025}. Analysis by \cite{bolinKeckGeminiPalomar2023} and \cite{markwardt2023} find the Neptune Trojan color distribution is distinct from the Jupiter Trojans but  similar to that observed in the trans-Neptunian region. Recent JWST mid-infrared spectroscopy of eight Neptune Trojans reveal similarities to dynamically excited trans-Neptunian objects (TNOs) and Centaurs which may indicate a shared origin where the populations have evolved due to experiencing different thermal environments \citep{Markwardt2025}. }

\update{Within their orbits, the Neptune Trojans preserve signatures of the mechanism that placed them into the 1:1 MMR with Neptune. Although the orbital elements do oscillate, the average orbital elements for long-lived Neptune Trojans do not change significantly on gigayear timescales \citep{Zhou2009,Zhou2011, 2011MNRAS.412..537L}. Therefore, the eccentricities, inclinations, and libration amplitudes of Neptune Trojan orbits provide insight and constraints on their origin whether trapped by a migrating Neptune \citep{2009AJ....137.5003N, 2009MNRAS.398.1715L,2010MNRAS.405.1375L,2015Icar..247..112P,2016A&A...592A.146G} or forming in situ with Neptune \citep{2009MNRAS.398.1715L,2011MNRAS.412..537L}. The Neptune Trojans have a broad inclination distribution, with many on dynamically excited, inclined, orbits. \citep{Sheppard2006, 2015Icar..247..112P,Lin2016,2019Icar..321..426L}.  \cite{Lin2021} find that the L4 Neptune Trojan size, inclination and eccentricity, distributions are better fit with a two component (population) model while the libration amplitude distribution can be fit with a single-component model. Whether this is because the Neptune Trojans originated from two distinct locations in the solar system or is a result of differing evolution at 30 au is not solved.  Whether this also applies to the L5 cloud remains an open question.}

\update{Most of what we know about the Neptune Trojans comes from the L4 cloud.} To date, over thirty Neptune Trojans have been discovered\footnote{\url{http://minorplanetcenter.org/iau/lists/NeptuneTrojans.html}}.  Until recently, only two objects were known to have gigayear stable orbits within Neptune’s trailing Lagrange point (L5) cloud \citep{2010Sci...329.1304S}. For over 15 years, the L5 cloud has been positioned on or close to the Galactic Center. Fields with $<$ 10$^{\circ}$ Galactic latitude have been challenging to search for moving solar system bodies due to the high stellar background and crowded fields. A large swath of the on-sky projected L5 Neptune Trojan cloud has now emerged from the Galactic plane. The Neptune Trojans clouds cover thousands of square degrees of sky, and the combination of the depth (8.4-m combined primary and tertiary mirror) and sky coverage (9.6 deg$^2$ \update{field of view [FOV]}) makes the NSF-DOE Vera C. Rubin Observatory \citep{2019ApJ...873..111I, 10.71929/rubin/2571927} an ideal discovery machine for this distant small body reservoir. 

In 2025, Rubin Observatory's Science Validation \citep[SV;][]{claver2025, guy2026} survey serendipitously observed fields that overlap with part of the L5 cloud.  In this Letter, we present the physical and orbital characterization of two new primordial L5 Neptune Trojans, 2025 MH348 and 2025 NN80, using the Rubin astrometry and photometry publicly reported to the MPC. These new discoveries double the inventory of known primordial L5 Neptune Trojans allowing for a more robust comparison of the trailing (L5) and leading (L4) clouds. We review the discovery circumstances of these two Neptune L5 Trojans within Rubin Observatory's SV survey and discuss their orbits and dynamical stability  within the L5 cloud. We present our analysis of broadband colors and phase curves utilizing  Rubin multiband photometry, and compare the characteristics of the L5 cloud to their counterparts at L4. We also include in our analysis 2025 MD138, a candidate L5 Neptune Trojan with a 38 day arc that was also found during  the Rubin SV survey. 

\update{This Letter, presents the first characterization of objects discovered by Rubin Observatory's moving object discovery pipelines using observations taken in an LSST-like cadence. With the start of the LSST, new discoveries are expected in real-time over the coming months. This work serves as an exemplar for the LSST-era. We provide the reader with Jupyter notebooks of our analysis; one for our color and phase curve calculations and one for our dynamical assessments on Zenodo at  \url{https://zenodo.org/records/22967949} \citep{schwamb_2026_22967949}.}

\section{Rubin Detection and Observations}

Rubin Observatory \citep{2019ApJ...873..111I} is located on Cerro Pach\'{o}n in Chile. It has been under construction since 2015. The first photon was collected in April 2025, and the milestone of first light was achieved in June 2025. The Observatory's primary mission is to complete the ten-year wide-field Legacy Survey of Space and Time \citep[LSST;][]{2009arXiv0912.0201L,2019ApJ...873..111I, 2022ApJS..258....1B} which is expected to begin observations in late June 2026. The  unique design of the 8.4-m Simonyi Survey Telescope combines the primary and tertiary mirror in one single continuous mirror surface to create an effective aperture of 6.49-m \citep{2024SPIE13094E..09S}. The LSST Camera \citep[LSSTCam; ][]{2024SPIE13096E..1SR, 10.71929/rubin/2571927} is located at prime focus. The 3.2 gigapixel camera spans 9.6 deg$^2$ over a 3.5 degree diameter \update{FOV}, with a delivered pixel scale of 0.2$^{\prime\prime}$ per pixel. The camera is equipped with 6 broadband filters ($ugrizy$) spanning the near-ultraviolet to near-IR (NIR). Further details of the LSST filter system can be found in  \citet{2019ApJ...873..111I} and  \cite{10.71929/rubin/2571927}. 

\subsection{Rubin Science Validation Survey}

\begin{figure}
\begin{center}
    \includegraphics[width=0.80\columnwidth]{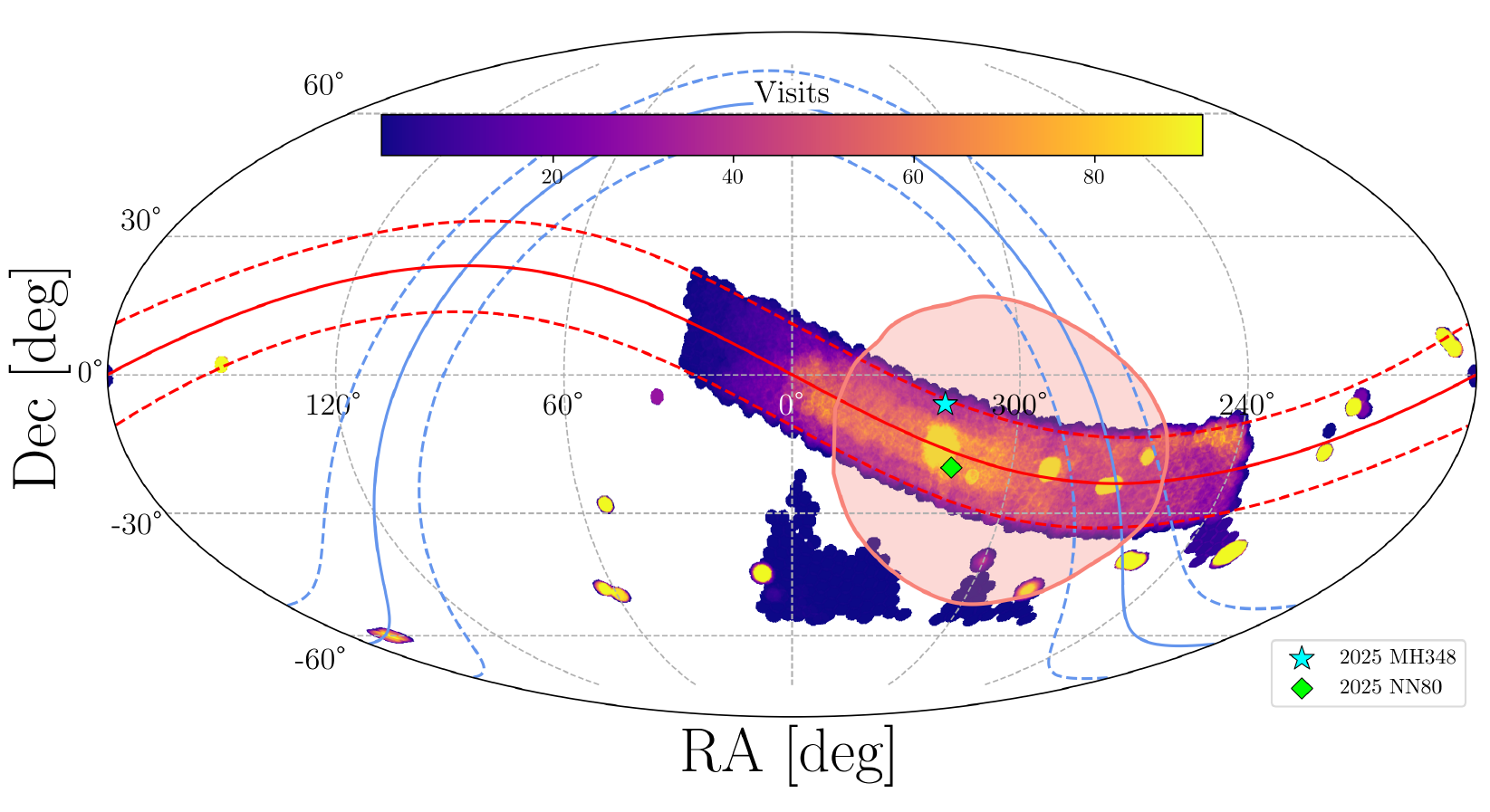}
     \caption{\label{fig:SV} Footprint of the SV observations, color-coded by number of visits. The ecliptic plane is indicated in red, and the Galactic plane is indicated in blue. The dashed red and blue lines represent $\pm$ 10$^{\circ}$ ecliptic and Galactic latitude, respectively. The location of the L5 Neptune Trojan cloud (derived from the model used in \citet{2026AJ....171...90M}) is indicated by the shaded orange area, with the locations of the two new stable L5 Trojans indicated.}
\end{center}
\end{figure}

The SV survey began on UT 2025 June 20 \citep{claver2025, guy2026}, operating in all six LSST $ugrizy$ bandpasses with 30~s exposures reaching a single-visit depth of $m_r \sim 24$~mag across a wide-field footprint concentrated primarily along $\pm$10$^{\circ}$ of the ecliptic plane; this encompassed $\sim$3000~deg$^2$ over the course of the commissioning window (see Figure \ref{fig:SV}). Designed to validate system performance and demonstrate operational readiness ahead of the 10-year LSST baseline, the SV footprint was imaged nightly, rather than at the $\sim$$3-4$ night cadence of the nominal wide-fast-deep strategy. This accelerated cadence was intended to accumulate on the order of a year's worth of equivalent survey depth within the commissioning window, whilst simultaneously exercising solar system detection and linking pipelines at LSST scale.

Whilst the SV survey was not designed with Neptune Trojan discovery as a specific objective, its ecliptic coverage serendipitously encompassed the on-sky position of the L5 cloud emerging from the galactic plane in 2025 (Figure~\ref{fig:SV}). This resulted in the detection of 2025 MH348 and 2025 MD138 on UT 2025 June 21, just one day after SV operations commenced, and 2025 NN80 on UT 2025 July 7; all observations are summarized in \update{Table~\ref{tab:obs}}. The image data associated with the SV survey are expected to be made available to data rights holders as part of Rubin Observatory's Data Preview 2 \update{\citep[DP2; ][]{DP2,DP2techreport}}. 

\subsection{Discovery, Astrometry, and Photometry}

As presented in \cite{2026ApJ...996L..33G}, LSSTCam observations are reduced using standard techniques via the Rubin Data Management (DM) System \citep{2017ASPC..512..279J,2022arXiv221113611O}. Flat fielding, bias subtraction, background subtraction, astrometric and photometric calibration,  source detection, astrometric measurements, and photometry were performed on the LSSTCam observations in a similar manner to that described in \cite{2026AJ....171..360V} for the LSST Commissioning Camera (LSSTComCam). A more detailed description of the software and algorithms that comprise the LSST Science Pipelines used within the Rubin DM System to reduce LSSTCam observations is described in \cite{PSTN-019}. 

2025 MH348, 2025 NN80, and 2025 MD138 were found in a preliminary search of the SV data performed using catalogs of sources identified on single-frame LSSTCam images, comparing sources detected in the same patch of sky across multiple nights to filter out static sources and thus identify transients and moving solar system objects. 
The Rubin solar system Processing framework \citep[SSP;][]{lsstSSP,2026AJ....171..360V} searched these catalogs using two techniques: an expanded version of the \texttt{HelioLinC} algorithm \citep{2018AJ....156..135H,2022DPS....5450404H} and a new code tuned to look for distant moving objects, \texttt{pointdexter} (Holman et al., in preparation). 

The detections and observations of 2025 MH348, 2025 NN80, and candidate 2025 MD138 were reported to the MPC in February 2026.  The classification of these newly designated small bodies as potential L5 Neptune Trojans was highlighted in discussions on the Minor Planet Mailing List (MPML; a forum for the amateur and professional astronomers within the minor planet community)\footnote{\url{https://groups.io/g/mpml/topic/neptune_trojan/117889052}}. We accessed the Rubin photometry through an export of the Asteroid Institute's real-time replica of the MPC PostgreSQL database\footnote{\update {Download links and tutorials are accessible via \url{https://b612.ai/rubin-mpc-downloads/} in comma-separated-value (CSV), Parquet, and SQLite formats. The data files are updated automatically when new Rubin observations (observatory code X05) are published by the MPC. The contents of these files have the same format as the observations table (obs\_sbn) within the MPC PostgreSQL database which is described at \url{https://data.minorplanetcenter.net/mpcops/documentation/obs-sbn/}}} \citep{koumjian_2025_17047589} and via the MPC Explorer\footnote{\update{\url{https://data.minorplanetcenter.net/explorer/}}} web interface. The Rubin photometry is presented in Table~\ref{tab:obs}. 

The detection efficiency of the preliminary Rubin moving object search has not yet been reported to the community. Therefore, we do not attempt to estimate a size distribution/population estimate from this sample or compare to the \cite{2026AJ....171...90M} predictions for the SV survey's Neptune Trojan discovery yield.  We analyze the orbits (the fitting of which includes additional astrometry linked since the discovery of these objects (see Section \ref{sec:orbits}) and average colors (Section \ref{sec:colors}) for all three objects. We also calculate a phase curve parameters for 2025 NN80 in Section \ref{sec:colors}. 

\section{Orbits and Dynamical Stability}
\label{sec:orbits}

\begin{figure}
    \includegraphics[width=0.50\columnwidth]{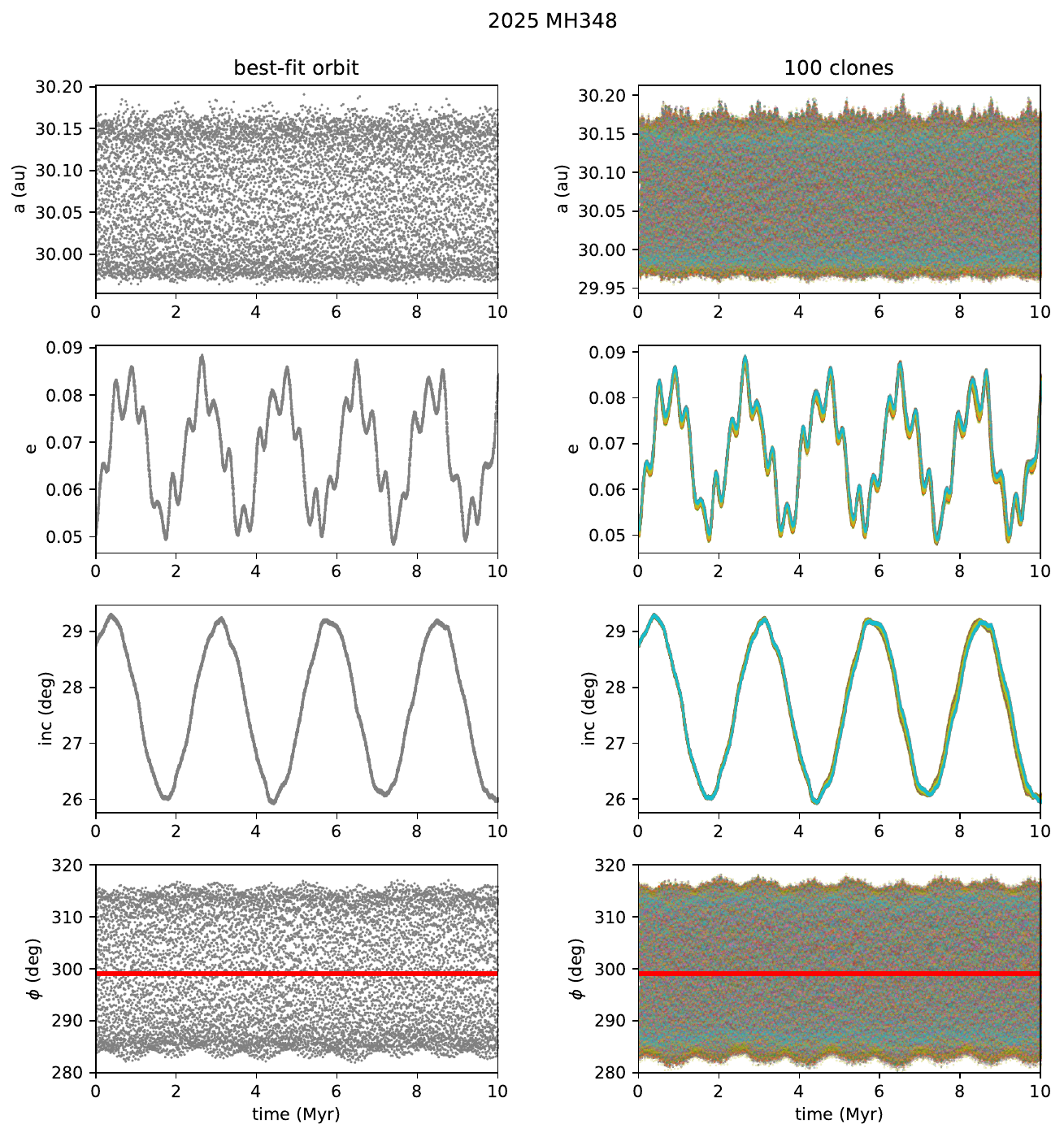}
      \includegraphics[width=0.50\columnwidth]{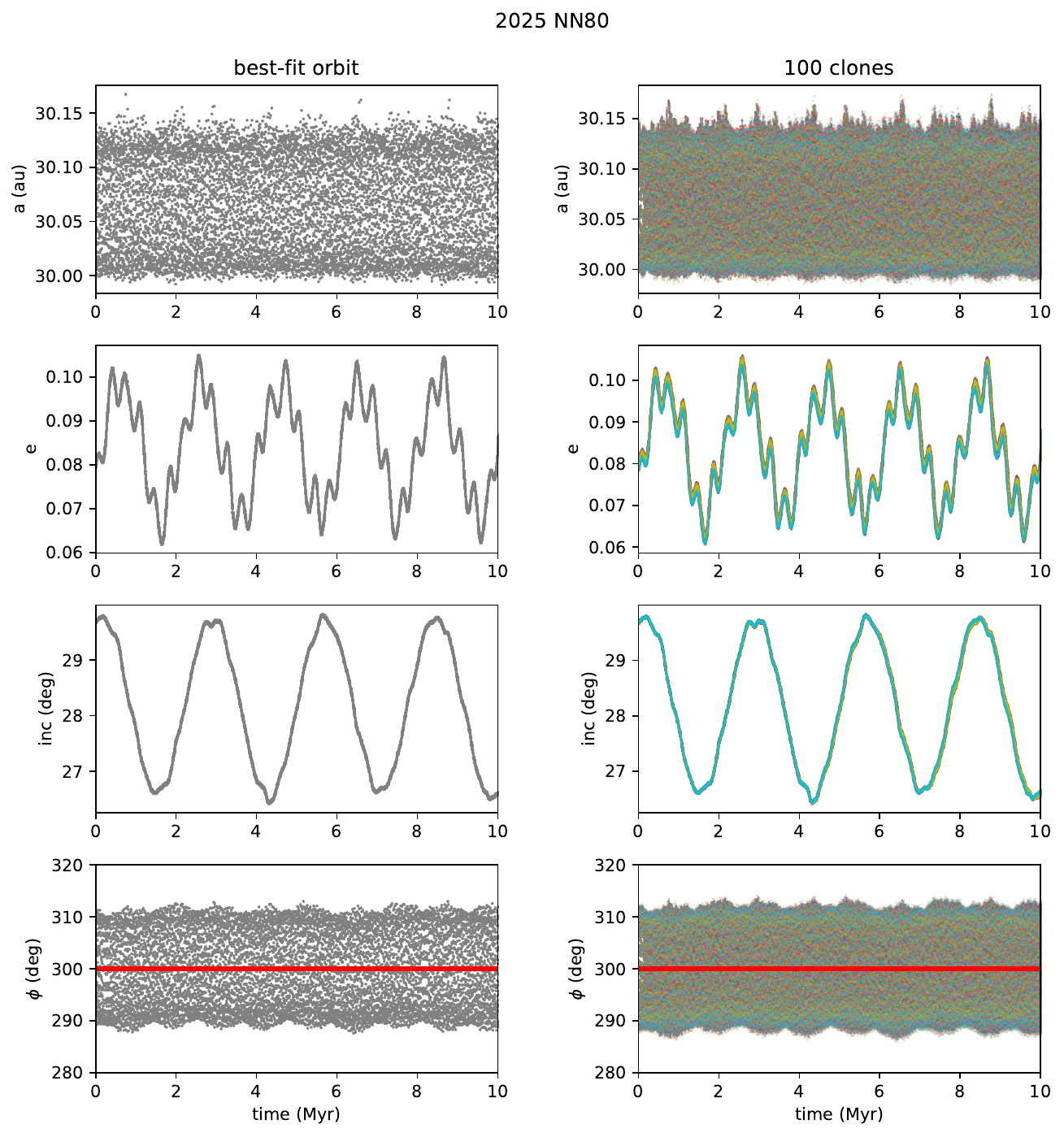}
     \caption{\label{fig:dynamics10Myr}
     The 10 Myr evolution of 2025 MH348 and 2025 NN80 in $a$, $e$, $i$, and $\phi=\lambda - \lambda_N$. The black points show the best-fit orbit and the colored points show each of the 100 clones sampled from the covariance matrix. \update{The red lines in the bottom panels indicate the libration center of the best-fit orbit (see Table~\ref{tab:evolution}).}}
\end{figure}

\begin{figure}

    \includegraphics[width=0.50\columnwidth]{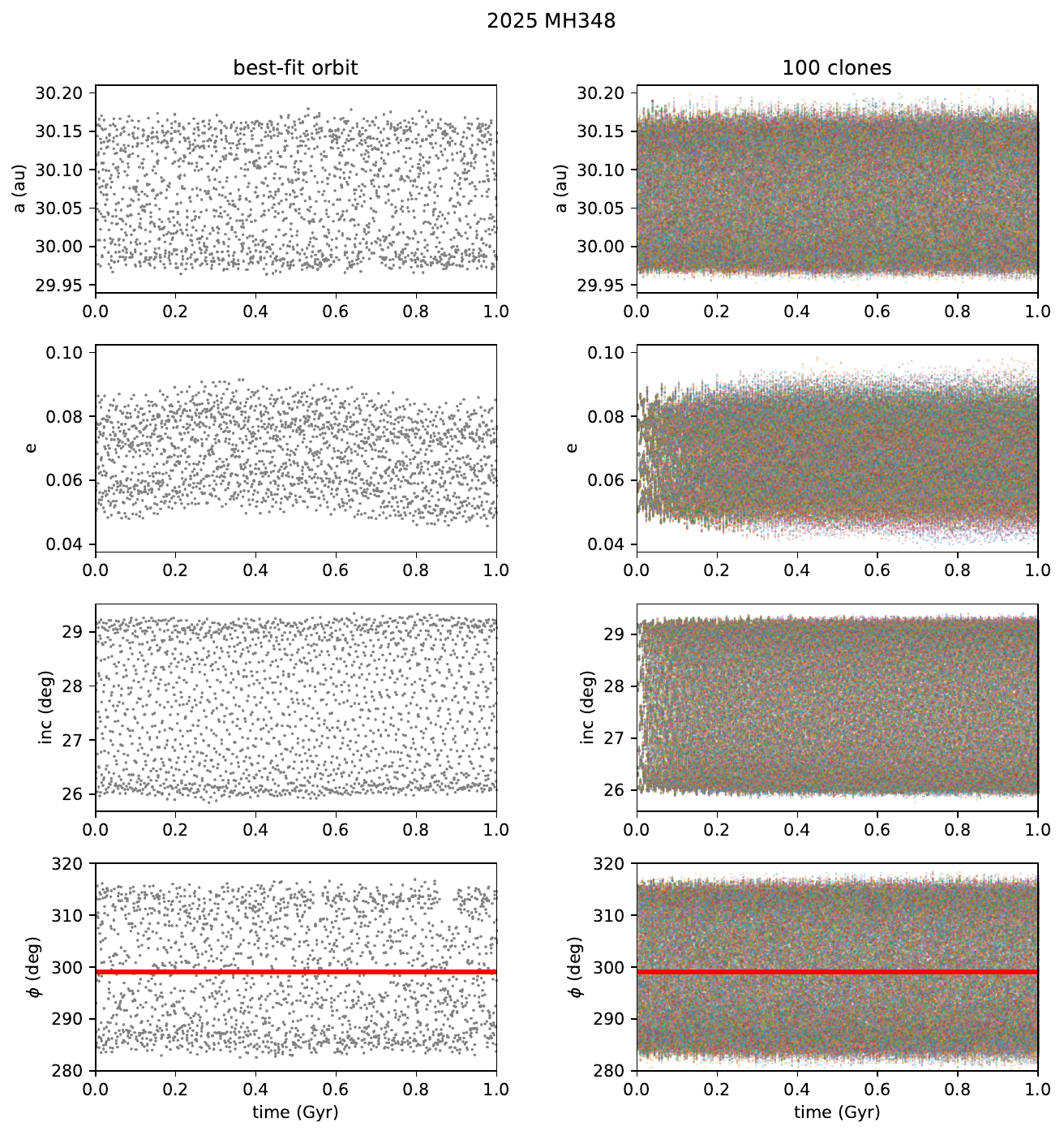}
      \includegraphics[width=0.50\columnwidth]{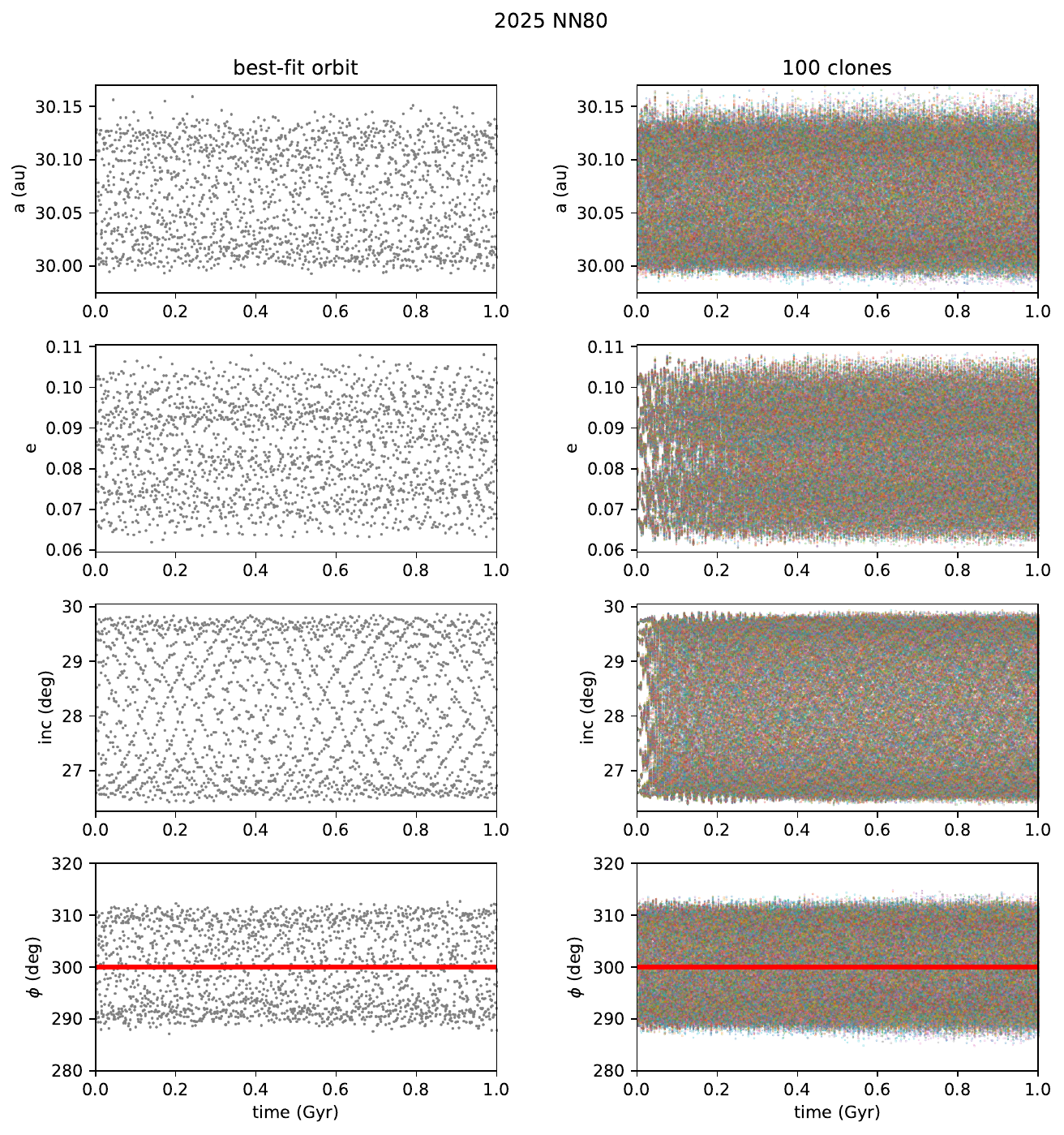}
     \caption{\label{fig:dynamics1Gyr}
     The 1 Gyr evolution of 2025 MH348 and 2025 NN80 in $a$, $e$, $i$, and $\phi=\lambda - \lambda_N$. The black points show the best-fit orbit and the colored points show each of the 100 clones sampled from the covariance matrix. \update{The red lines in the bottom panels indicate the libration center of the best-fit orbit (see Table~\ref{tab:evolution}).}}
\end{figure}

 At the time of writing, previous Pan-STARRS1 \cite[Panoramic Survey Telescope and Rapid Response System 1;][]{2016arXiv161205560C} observations have now been associated with 2025 MH348 and 2025 NN80, sufficiently extending the observational arcs for multi-opposition orbit fits; 2025 MD138 has only the 38 day arc from the SV survey, which accounts for its much larger orbital uncertainties. 
 Both 2025 MH348 and 2025 NN80 are stable L5 Neptune Trojans on Gyr timescales based on integrations of their JPL orbit fits and 100 clones sampling the JPL orbit fit covariance matrix. 
We used the Small Body Dynamics Tool (\texttt{SBDynT}) for the integrations of all three objects \citep{Spencer:2026} which are performed using the Mercurius integrator in \textsc{rebound} \citep{rebound,reboundmercurius} and include perturbations from the Sun and the four giant planets.
Table~\ref{tab:orbits} lists the current best-fit orbital elements for the three objects we consider and the 1-$\sigma$ uncertainties from the JPL Small-Body Database\footnote{\url{https://ssd.jpl.nasa.gov/tools/sbdb_lookup.html}}. 
We performed high-resolution 10 Myr integrations (Figure~\ref{fig:dynamics10Myr}) to measure libration parameters for 2025 MH348 and 2025 NN80, which are given in Table~\ref{tab:evolution} (we also list the libration parameters of the other two stable L5 Trojans for comparison, determined from \texttt{SBDynT} integrations of the JPL orbits). We then extended these integrations to 1 Gyr (Figure~\ref{fig:dynamics1Gyr}), confirming that all clones of 2025 MH348 and 2025 NN80 remain in Neptune's 1:1 resonance; at t=1 Gyr, their orbital parameters remain nearly identical to the first 10 Myr values listed in Table~\ref{tab:evolution}. \update{Only the best-fit orbit and 36 of the 100  clones of 2025 MD138 librate in the 10 Myr integrations (with 23 surviving as Trojans at 1 Gyr), therefore we do not characterize its libration parameters.} \update{We note that for all three objects, we also ran the \texttt{SBDynT} machine learning orbit classifier \citep{2025mlsm.book..173V} on all the clone integrations, which generated the same classifications as our own analysis. We provide an example notebook of our \texttt{SBDynT} analysis on Zenodo at \url{https://zenodo.org/records/22967949} \citep{schwamb_2026_22967949}.}

\begin{figure}
\begin{center}
      \includegraphics[width=0.50\columnwidth]{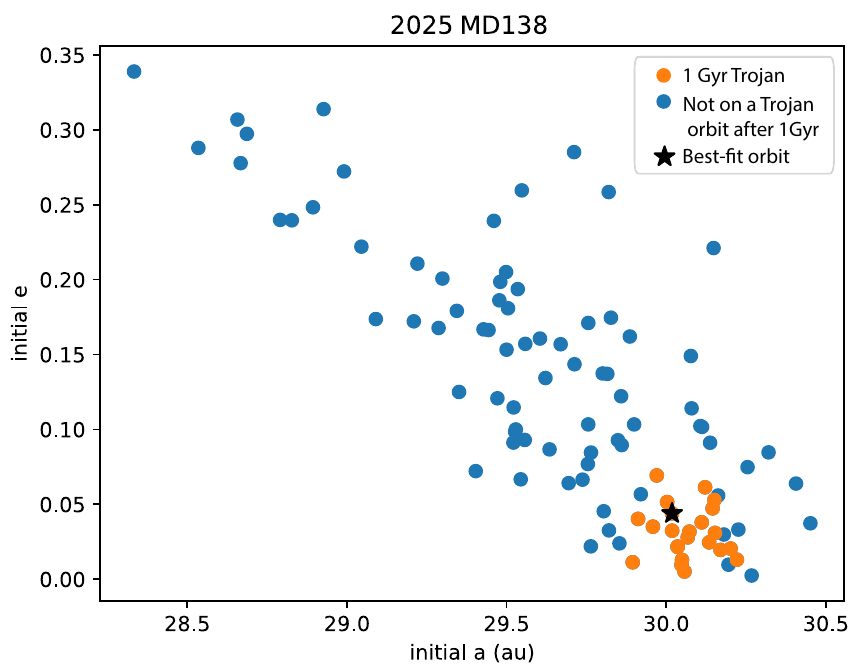}
     \caption{\label{fig:2025MD138}
     The long-term stability of 100 clones of 2025 MD138 sampled from its JPL covariance matrix. The initial $a$ and $e$ of the clones are plotted as circles with the color denoting the state at the end of the 1 Gyr $n$-body integrations. The black star denotes the current best-fit orbit, which is stable at 1 Gyr.}
\end{center}
\end{figure}

It is of note that all four known stable L5 Trojans have inclinations exceeding 25$^\circ$. 
This likely reflects both the dynamically hot nature of the Neptune Trojans as a whole \citep{Lin2021, Sheppard2006} and the inclination-dependent stability of the resonance.
In particular, \citet{Dvorak2007, Zhou2009, Zhou2011} modeled the long-term stability of the Neptune Trojan phase space and demonstrated that orbital stability exhibits a dependence on inclination, peaking near $30^\circ$ and below $10^\circ$. 
The four known L5 Trojans at $\sim 30^\circ$ populate one of these prominent stability islands. 
This clustering of L5 Trojans at high inclinations ($i > 25^\circ$) highlights an intriguing comparison with the L4 population. 
As demonstrated by \citet{Lin2016, Lin2021}, the L4 inclination distribution exhibits a peculiar enhancement in the low-inclination regime, which appears to be exclusively confined to larger Neptune Trojans. 
Given that our current sample of definitive stable L5 Trojans consists of only four objects, it remains premature to draw definitive conclusions regarding their overall size-inclination distribution. 
We note that 2025 MD138 is a \textit{potentially} stable L5 Trojan, and it has an inclination of $i=6^\circ$.
Figure~\ref{fig:2025MD138} shows the distribution of clones of 2025 MD138's orbit;  while the best-fit orbit is consistent with Gyr stability around the L5 libration point, only 23 of 100 clones sampling the orbit fit uncertainties are also stable. 
Both its semimajor axis and its eccentricity have large uncertainties due to its only 38 day arc; 
additional follow-up observations will be necessary to refine its orbit and test whether it is indeed a low-inclination stable L5 Trojan. We note that all 23 surviving clones (and 2025 MD138's best-fit orbit) retain their low inclinations at the end of the 1 gigyear $n$-body integrations. 
The presence or absence of low-inclination L5 counterparts  warrants further investigation as the trailing population sample grows. 

\begin{deluxetable}{c c c c c c c c c c}
    \tabletypesize{\scriptsize}
    \tablecaption{Orbit fits used in our integrations \label{tab:orbits}}
    \tablewidth{\columnwidth}
    \tablehead{\colhead{object}  & \colhead{$e$} & \colhead{$q$ (au)} & \colhead{$i$ ($^{\circ}$)} & \colhead{$\Omega$ ($^{\circ}$)} & \colhead{$\omega$ ($^{\circ}$)} & \colhead{$t_p$ (JD)} & \colhead{$\lambda - \lambda_N$ ($^{\circ}$)} &  \colhead{epoch (JD)} & \colhead{data arc (day)}}
    \startdata
    2025 MH348 & 0.05130  & 28.6299 & 28.70903  & 154.87848 & 82.432  & 2448503.73 & 310.07 & 2459592.5 & 2595\\
    (1-$\sigma$) & 0.00035 & 0.0071 & 0.00020 & 0.00024 &  0.078 &  9.86 & 0.03 & &\\
    \hline   
    2025 NN80 & 0.07841 & 27.755  &  29.68792 & 126.21271 & 142.36  & 2454476.86 & 305.78 & 2460538.5 & 1547\\
    (1-$\sigma$) &0.00039  & 0.014 & 0.00050 & 0.00011 &  0.46 &  63.20 & 0.08 & &\\
    \hline   
    2025 MD138 & 0.045 & 28.6  &  6.044 & 220.1 & 164  & 2472639 & 312 & 2460871.5 & 38\\
    (1-$\sigma$) & 0.12  & 4.0 & 0.0002 & 0.3 &  76 &  10028 & 9 & & \\
    \enddata
    \tablenotetext{}{Orbit fits and covariance matrices taken from JPL's Small Body Database Browser. The solution date for 2025 MH348 is 2026-06-05 02:26:14, for 2025 NN80 is 2026-06-05 02:30:02, and for 2025 MD138 is 2026-06-05 02:16:37. \update{We also include each object's mean longitude relative to Neptune ($\lambda - \lambda_N$) at the orbit-fit epoch along with the 1-$\sigma$ uncertainties from our cloning procedure (1-$\sigma$ uncertainties for the other parameters are directly from JPL).}
    } 
\end{deluxetable}

\begin{deluxetable}{l c c c c }
    \tabletypesize{\scriptsize}
    \tablecaption{10 Myr dynamical evolution parameters of the known stable L5 Neptune Trojans \label{tab:evolution}}
    \tablewidth{\columnwidth}
    \tablehead{\colhead{object}  & \colhead{$<\lambda-\lambda_N>$ ($^\circ$)} &  \colhead{$\Delta(\lambda-\lambda_N)$ ($^\circ$)} & \colhead{$<e>$} & \colhead{$<i>$ ($^\circ$)}}
    \startdata
    2025 MH348 & $299.06\pm0.15$  & $33\pm1$ & $0.06816\pm0.0003$  & $27.593\pm0.002$  \\
    2025 NN80 & $300.06\pm0.08$  & $23.2\pm0.9$ & $0.08366\pm0.0005$  & $28.1050\pm0.0007$  \\    
    2008 LC18 & $298.09\pm0.37$  & $37\pm2$ & $0.08150\pm0.0001$  & $25.943\pm0.002$  \\    
    2011 HM102 & $300.03\pm0.07$  & $22.4\pm0.8$ & $0.087294\pm0.00003$  & $27.735\pm0.001$  \\   
    \enddata
    \tablenotetext{}{Libration center ($<\lambda-\lambda_N>$) and amplitude ($\Delta(\lambda-\lambda_N)$) and average eccentricity and inclination for the 4 stable L5 Trojans based on 10 Myr integrations of their observed orbits. Nominal values are for the best-fit orbit and uncertainties are the standard deviation amongst the 100 clones.} 
\end{deluxetable}

\section{Colors, Phase Curves, and Predicted Surface Composition}
\label{sec:colors}

\begin{deluxetable}{l c c c c c c c c c c c}
\tabletypesize{\tiny}
\tablecaption{Number of Rubin observations per filter, mean photometric colors, and phase curve fits \label{tab:colors}}
\tablewidth{\columnwidth}
\tablehead{\colhead{object}  & \colhead{\#$g$} & \colhead{\#$r$} & \colhead{\#$i$} & \colhead{\#$z$} & \colhead{($g$-$r$)} & \colhead{($r$-$i$)} & \colhead{($r$-$z$)}  & \colhead{$H_r$} & \colhead{$\beta_r$} & \colhead{$H_i$} & \colhead{$\beta_i$} }
\startdata
2025 MH348 & 3 & 5 & 3 &  0 & 0.46 $\pm$ 0.08 & 0.45 $\pm$ 0.09 & - &  - & - & - & -  \\
2025 NN80 & 5 & 7 & 12 &  8 & 0.47 $\pm$ 0.05 & 0.30 $\pm$ 0.04 &0.42 $\pm$ 0.05 &  7.71 $\pm$  0.05 & 0.253 $\pm$ 0.055 & 7.54 $\pm$  0.03 & 0.182  $\pm$ 0.050  \\
2025 MD138\tablenotemark{a} & 2 & 6 & 6 &  1 & 0.37 $\pm$ 0.06 & 0.39 $\pm$ 0.05 & -& - & - & - & - \\
\enddata
 \tablenotetext{a}{Candidate long-term stable Neptune Trojan.}
\tablenotetext{}{All filters, estimated colors, and phase function parameters are reported in the LSST photometric system.}
\end{deluxetable}

\begin{figure}
\begin{center}
      \includegraphics[width=0.45\columnwidth]{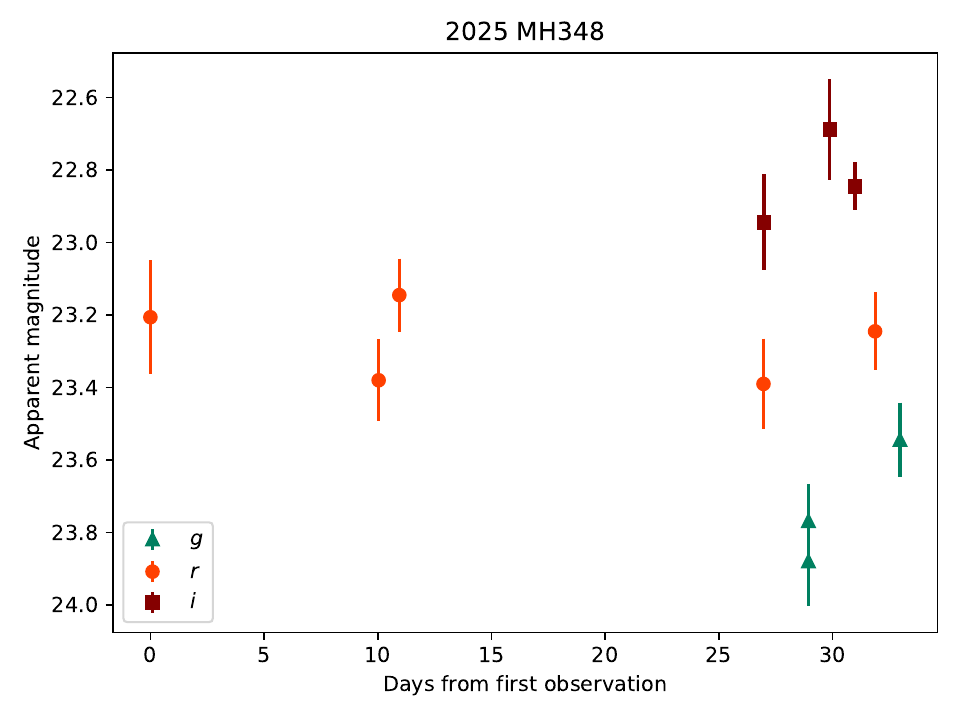}
       \includegraphics[width=0.45\columnwidth]{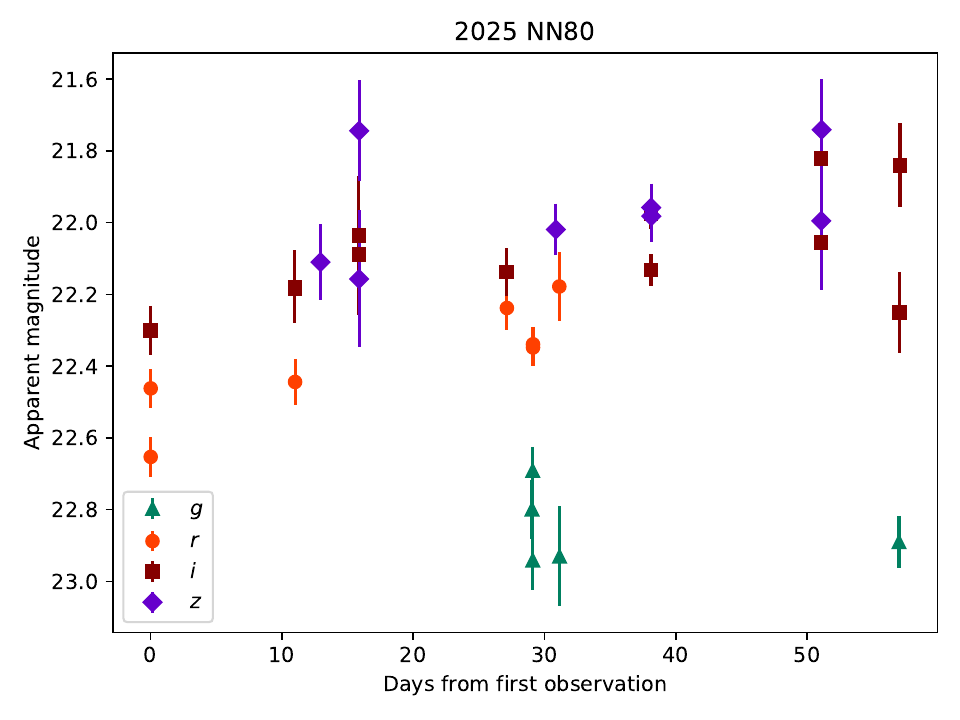}
    \includegraphics[width=0.45\columnwidth]{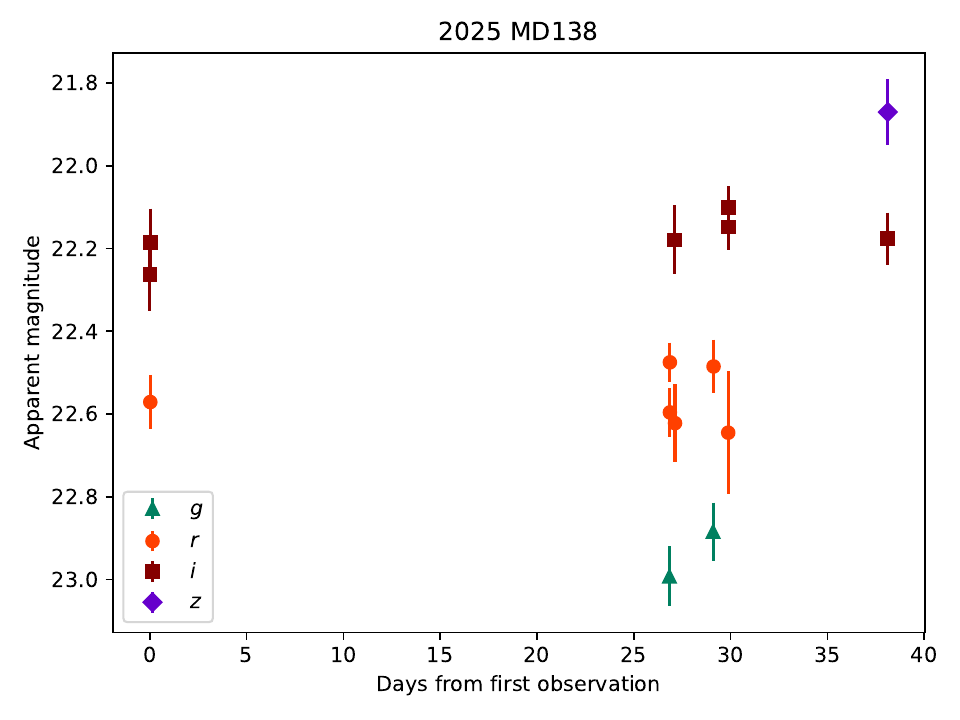}
     \includegraphics[width=0.45\columnwidth]{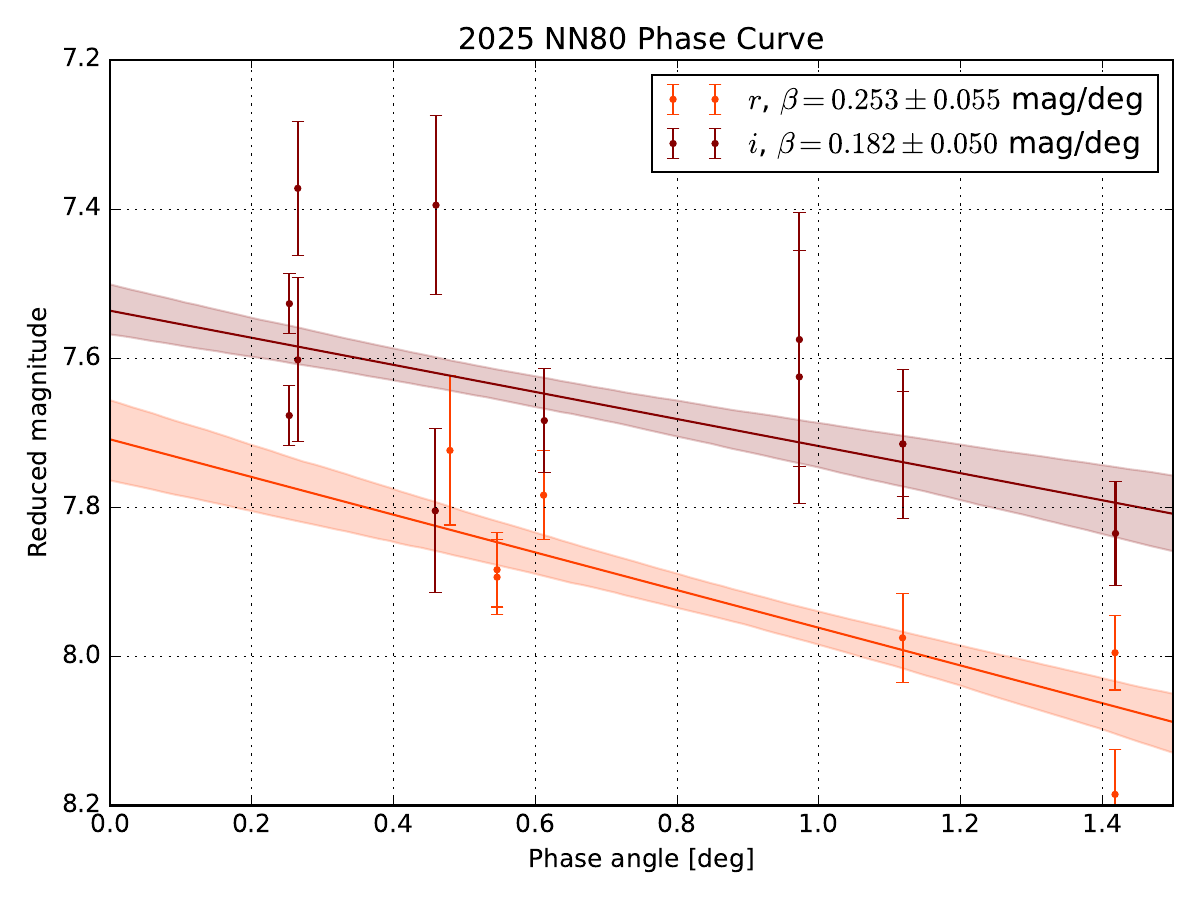}
\end{center}
    \caption{\label{fig:lcs_phase} Light curves (apparent magnitude per filter as a function of time) from the Rubin multi-band photometry are displayed in the top row and bottom left plots. Bottom right: the reduced magnitude measurements for 2025 NN80 in the $r$ and $i$ band as a function of phase angle. The solid curve represents the best-fit phase curve for each band and the shaded region the $1\sigma$ contours.}
\end{figure}

Light curves for 2025 MH348, 2025 NN80, and  2025 MD138 are shown in Figure \ref{fig:lcs_phase}.  The data are too sparsely separated in time for full rotational light-curve analysis, but we note that 2025~MD${138}$ shows very little variation. 2025~NN${80}$ and 2025~MH${348}$ show larger variation, but within the error bars do not appear to have inordinately high amplitudes. Variation trend by color are self-consistent, and we interpret this to mean that the objects are not significantly color-variable and suggest that within the photometric measurement uncertainty the surface composition is consistent across the whole of each object. Given the small number of data points per filter per object, as summarized in Table \ref{tab:colors} and the small phase angle range ($<$ 2$^{\circ}$), we calculate mean apparent magnitudes per band.

\begin{figure}
\begin{center}
      \includegraphics[width=0.49\columnwidth]{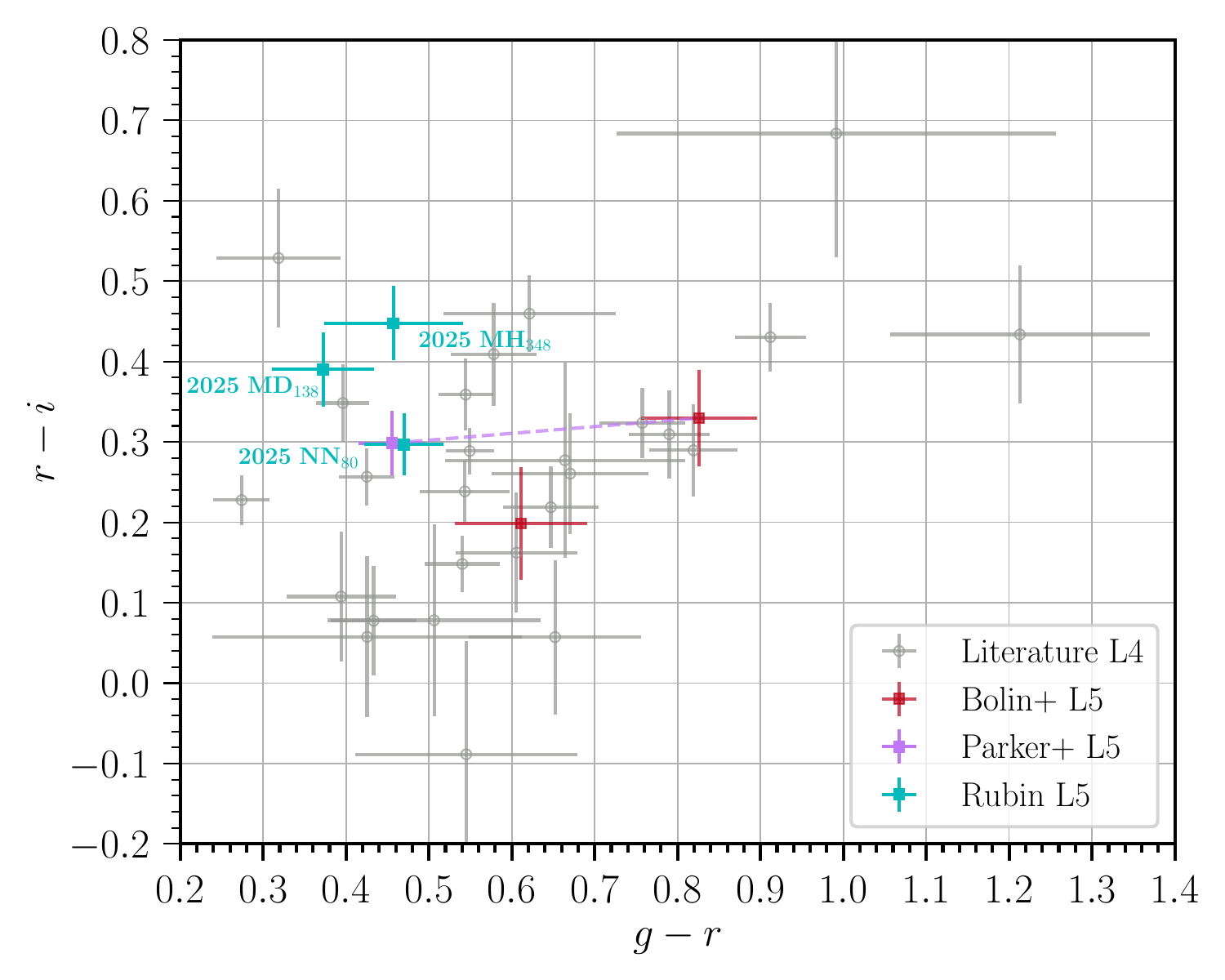}
      \includegraphics[width=0.49\columnwidth]{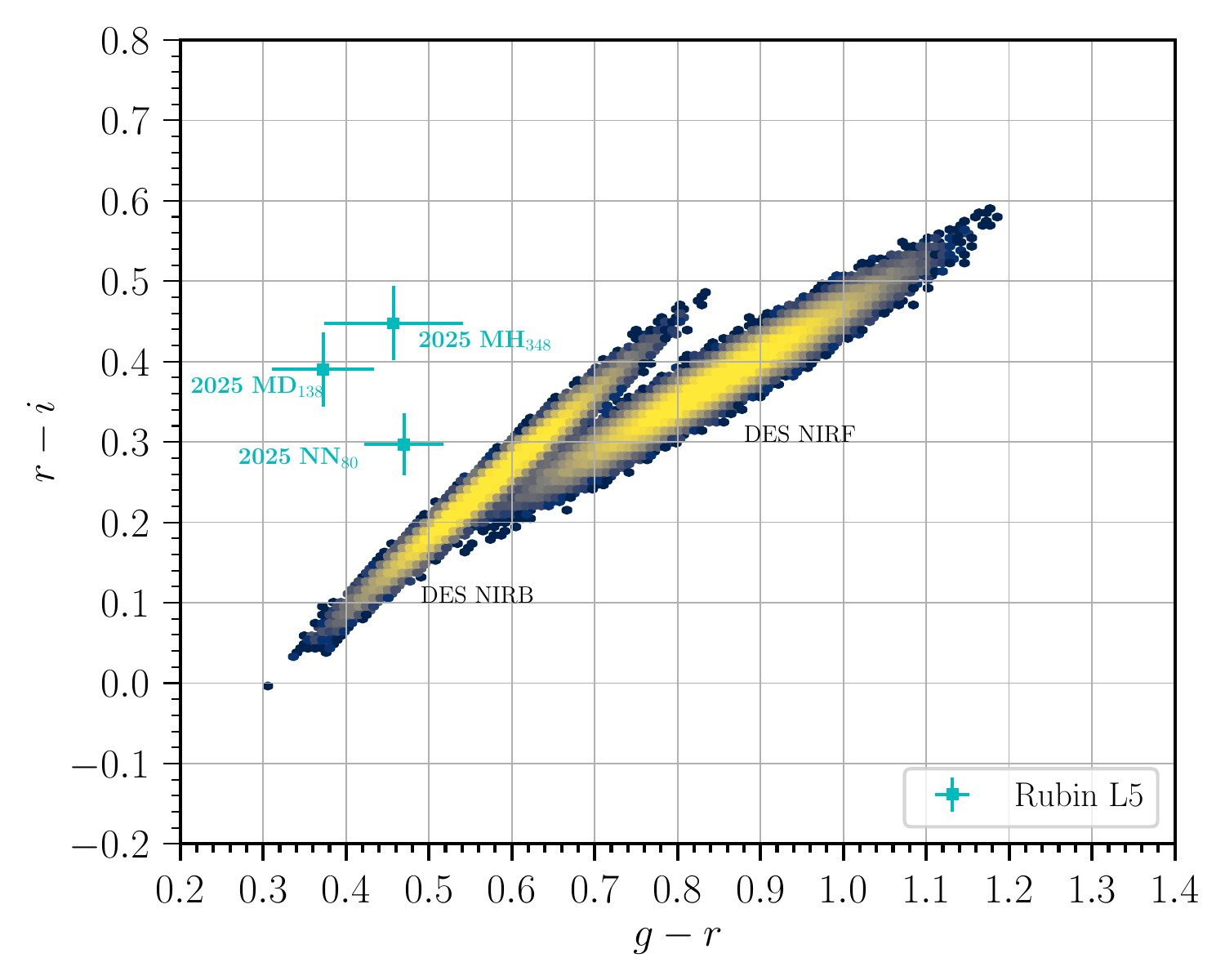}
     \caption{\label{fig:colors} \emph{Left:} $g-r$ and $r-i$ colors of the Rubin L5 Neptune Trojans (in teal), as well as the objects from \cite{markwardt2023} and \cite{bernardinelli2023} (in gray). The  L5 Trojans from \cite{bolinKeckGeminiPalomar2023} are shown in red, and the L5 Trojan from \cite{parker2013} is shown in purple. Colors for 2011 HM102 are reported in \cite{parker2013} and is also in the \cite{bolinKeckGeminiPalomar2023} sample, the two differing color measurements are connected by a \update{purple} dashed line. \emph{Right:} The same Rubin measurements, and a heat map showing samples from the two component Gaussian model of \cite{bernardinelli2025}.} 
\end{center}
\end{figure}

With these in hand, we compute ($g$-$r$) and ($r$-$i$) colors for all three objects and ($r$-$z$) for 2025 NN80  that can be compared to previous measurements of Neptune Trojans both in the L4 and the 2 previously known long-term stable L5 Trojans with published colors.  We present these results in Table \ref{tab:colors}. We compile colors from the literature, with  \cite{markwardt2023}, \cite{bolinKeckGeminiPalomar2023} and \cite{bernardinelli2023} being the largest compiled data sets of Neptunian Trojans, as well as one L5 Trojan from \cite{parker2013}. Since these colors are in different photometric systems, we use the conversions from \cite{RTN-099} to bring these colors into the LSST filter system\update{, presented in Appendix \ref{app:photometric}}. We present these comparisons in the left panel of Figure \ref{fig:colors}. It is clear that the L4 and the joint set of L5 from this work and from \cite{bolinKeckGeminiPalomar2023} occupy a similar color locus, with these new three objects having bluer colors (($g-r$) near $0.4$) than the two from \cite{bolinKeckGeminiPalomar2023}, and consistent with the object measured by \cite{parker2013}.

We can also try to interpret these colors in the context of the two-component color model of \cite{bernardinelli2025}, which uses data from the Dark Energy Survey (DES) to attempt to model the colors of TNOs as a combination of two Gaussian components, called near-IR bright (NIRB) and near-IR faint (NIRF), where the NIRB component is bluer and more dominant in the dynamically excited TNO classes. The right panel of Figure \ref{fig:colors} shows this model (in the LSST photometric system) in addition to the 3 objects from this study. Under the (reasonable) assumption of Gaussian distributed errors, we can formally assign a distance from each object to the closest component in the model (in this case, the NIRB component) by computing the Mahalanobis distance between object and model. Considering color $\mathbf{c} \equiv (g-r, r-i)$, we have that
\begin{equation}
    \mathcal{M} \equiv  \sqrt{\left(\Delta \mathbf{c}^\top  (\boldsymbol{\Sigma}_\mathrm{object} + \boldsymbol{\Sigma}_\mathrm{model})^{-1}  \Delta \mathbf{c}\right)},
\end{equation}
where $\Delta \mathbf{c} \equiv \mathbf{c}_\mathrm{object} - \mathbf{c}_\mathrm{model}$ and $\boldsymbol{\Sigma}_{\mathrm{object},\mathrm{model}}$ are the covariance matrices for object and model, respectively. Intuitively, we can interpret this distance as the multi-dimensional generalization of ``how many $\sigma$ away from the reference value'' each measurement is. When compared to the NIRB distribution, we have that 2025 MH348 and 2025 MD138 have $\mathcal{M} = 0.36$ and $\mathcal{M} = 0.51 $, respectively, being consistent with being sampled from the model. Despite being closer to the NIRB distribution, 2025 NN80's lower uncertainty leads to  $\mathcal{M} = 2.59$, which corresponds, roughly to a $p$-value $\lesssim 0.05$. 
If we include the measured $r-z$ as well, $\mathcal{M} = 2.25$, thus making this object consistent with the NIRB distribution. Since this is only one object (in a sample of three) and this is a low significance discrepancy only in two of the three measured colors, we refrain from interpreting this as a true outlier from the model, and we encourage that such statistical analysis like the one presented here be repeated with larger samples of objects coming from full LSST operations to determine whether there are differences between the Neptunian Trojan and TNO color distributions.

Only 2025 NN80 has small enough  error bars on the reported photometry and sufficient sampling, albeit sparse, across the observed  phase angle range to attempt an analysis of the phase function. Given the small number of measurements, this is only possible for the $r$ and $i$ bands. Reduced magnitudes, to remove distance effects,  were calculated and a linear phase function was fit independently for each filter. 
We show these measurements in Figure \ref{fig:lcs_phase} and the best-fit linear slope ($\beta$) and absolute magnitude ($H$) in Table \ref{tab:colors}. We note that the resulting $\beta_r$ and $\beta_i$ are within $1\sigma$ of each other and are consistent with those measured for TNOs, but more consistent with medium-large sized TNOs than small ones \citep{2007AJ....133...26R, 2002Icar..160...52S, 2016A&A...586A.155A,2018MNRAS.481.1848A, alvarez-candal2019, 2025MNRAS.540..460F,Verbiscer2019AJ,Verbiscer2022PSJ}. 

\begin{figure}
    \centering
    \includegraphics[width=0.70\columnwidth]{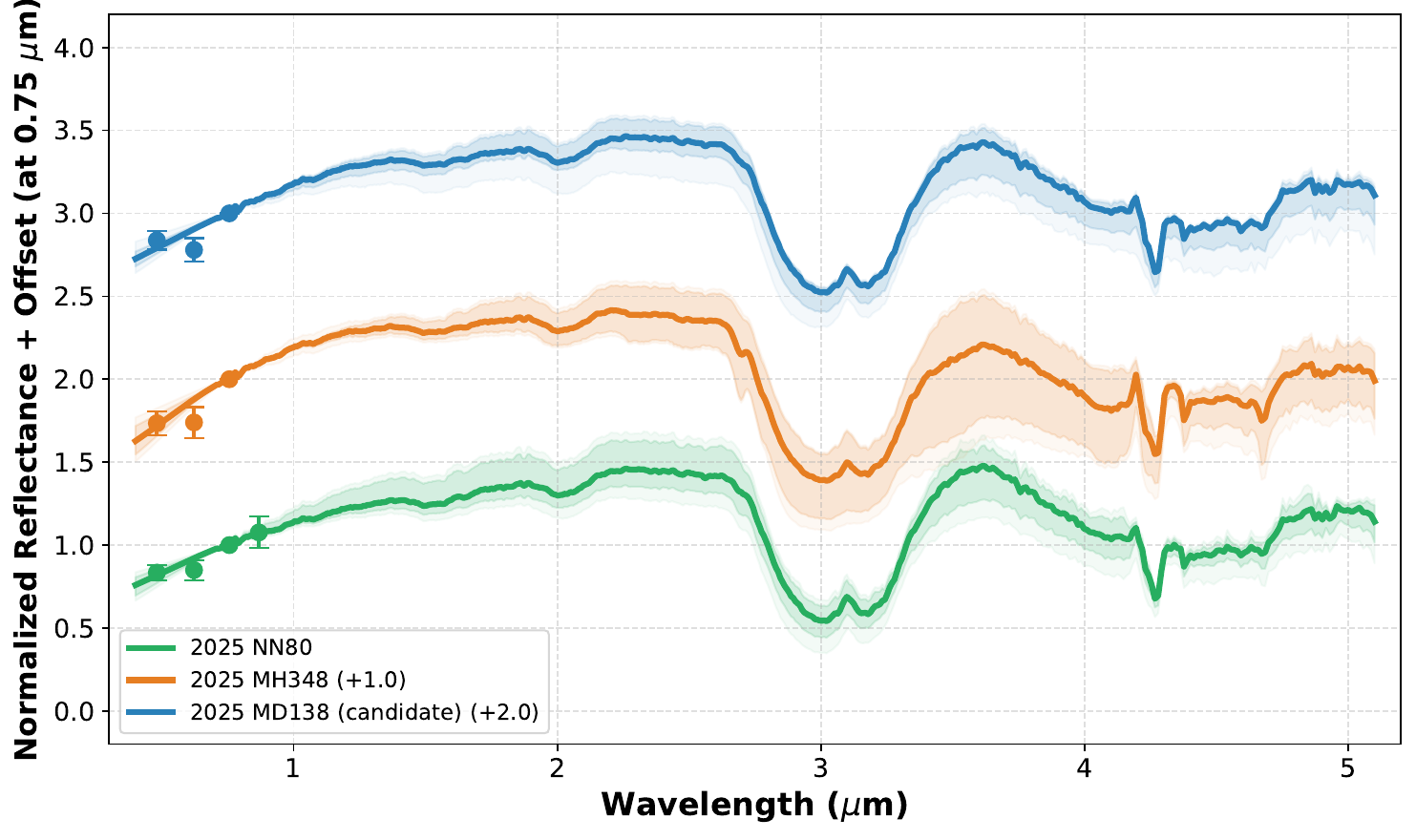}
     \caption{\label{fig:recon_spec}Posterior predictive spectra inferred from the input $gri$ or $griz$ photometry, where circular markers and vertical error bars denote the input values and propagated uncertainties (normalized at the $i$-band) used for the inference. Solid curves show the posterior median reconstruction, and darker/lighter shaded regions show the 68/90\% predictive interval.}
\end{figure}

We interpret the measured $gri$ or $griz$ colors using an optical extension of the probabilistic photometry-to-spectrum reconstruction framework developed for sparse NIR photometry by \update{\citet{Lin2026, lin2026_color_to_spec}}, obtaining posterior predictive optical–NIR spectra (Figure~\ref{fig:recon_spec}). The reconstructions suggest that the three objects are most consistent with the water-type spectra \citep{Pinilla-Alonso2025, Holler2025}, which is common within the Neptune Trojan population \citep{Markwardt2025}. However, we caution that the detailed spectral morphologies remain weakly constrained by optical colors alone, particularly at longer wavelengths and should not be over-interpreted; this result should be treated as a probabilistic branch assignment rather than a substitute for NIR photometry or spectroscopy. We also note that, although 2025~MD${138}$ and 2025~NN${80}$ have different individual ($g-r$) and ($r-i$) colors, their combined ($g-i$) colors are nearly identical. Because the present $gri$-based reconstruction is primarily sensitive to the overall optical slope rather than to detailed curvature within the optical continuum, these two objects are mapped to similar regions of the spectral manifold. This likely explains their highly similar posterior spectral reconstructions, but should not be interpreted as evidence that their detailed NIR spectral morphologies are necessarily identical.

\section{Summary}

Long-term stable L5 Neptune Trojans 2025 MH348 and 2025 NN80 and candidate 2025 MD138 represent Rubin Observatory's first glimpse at Neptune's trailing Trojan swarm. We find that our estimated surface colors of these new objects combined with those previously reported for the other two stable L5 Neptune Trojans provide a sample that is consistent with the L4 Neptune Trojans. This further supports the argument that the long-term stable Neptune Trojans originated from the same parent population. Compositional studies of the Jupiter Trojans reveal similar conclusions for the leading and trailing swarms about Jupiter. For comparison, observations of the Jupiter Trojan L4 and L5 asteroid populations have found the L4 cloud has a larger range of colors than what is seen in the L5 cloud, but this been ascribed to the presence of asteroid families in the L4 cloud with less-red parent bodies \citep{fornasier2025spectrophotometry}. When the samples are bias-corrected, the two clouds are consistent with one another in terms of color properties as a whole \citep{2022PSJ.....3..269P,fornasier2025spectrophotometry, yoshida2026color}. 



The L5 cloud has historically been understudied due to the very small number of discoveries. \cite{2026AJ....171...90M} predict that the LSST will quadruple the total number of known Neptune Trojans and the number with well measured broadband colors and multi-filter phase curves. Tens to hundreds of LSST small body discoveries are expected to be found in the trailing swarm \citep{2026AJ....171...90M}. Future Rubin detections will be able to further test the symmetry of the L4 and L5 Neptune Trojans and probe whether the size-inclination distribution within the L5 cloud matches what \citet{Lin2016, Lin2021} have reported for the leading L4 swarm. \cite{2026AJ....171...90M} predict that most discoveries will occur during the first two years of the LSST. 

Sixteen years after the discovery of the first L5 Neptune Trojan \citep{2010Sci...329.1304S}, there are now four confirmed members. \update{The LSST began at the end of  2026 June.  Rubin survey operations were paused in mid-July after significant rain and snow  on Cerro Pach\'{o}n and the surrounding areas. Science operations are expected to begin again in early southern spring after a planned engineering shutdown. The LSST footprint encompasses nearly all of the sky expected to be populated with L5 Neptune Trojans. Fields overlapping the L5 cloud will be observed by Rubin Observatory in the next few months and over the next twelve months}. With the L5 cloud moving out of the Galactic center and the dawn of the LSST, we will soon get the best view of this small body reservoir. 2025 MH348, 2025 NN80, and candidate 2025 MD138 represent the tip of the iceberg for the Neptune Trojan science that will be enabled over the coming months and next several years by the Rubin Observatory.

\begin{acknowledgments}

M.E.S. acknowledges support in part from UK Science and Technology Facilities Council (STFC) grant ST/X001253/1. MES acknowledges travel support provided by STFC for UK participation in LSST through grant ST/X001334/1.
K.V. acknowledges support from NASA grants 80NSSC23K0886, 80NSSC23K1169,80NSSC23K0680, and 80NSSC26K1031. 
J.M. and C.O.C. acknowledge support from the DiRAC (Data-intensive Research in Astrophysics and Cosmology)Institute in the Department of Astronomy at the University of Washington. The DiRAC Institute is supported through generous gifts from the Charles and Lisa Simonyi Fund for Arts and Sciences, Janet and Lloyd Frink, and the Washington Research Foundation. 
H.W.L. acknowledges the support of the National Science Foundation under Grant No. AST‑2406527. 
C.O.C. and M.W. acknowledges support through Schmidt Sciences LLC. C.O.C. gratefully acknowledges support from the NASA CSSFP (grant No. 80NSSC26K0380). 
M.J. acknowledges  support from the University of Washington College of Arts and Sciences, Department of Astronomy, and the DiRAC Institute.
The work of S.G. is supported by NOIRLab, which is
managed by the Association of Universities for Research in
Astronomy (AURA) under a cooperative agreement with the
US National Science Foundation.
A. W. was supported by UK STFC grant UKRI1776. 
R.R.L. was supported by the UK STFC grant ST/V506990/1.

We thank Breakthrough Listen, the Astrophysics Research Centre at Queen's University Belfast (QUB), and the QUB  School of Mathematics and Physics for their generous support of the LSST solar system First Data Sprint. Breakthrough Listen is managed by the Breakthrough Initiatives, sponsored by the Breakthrough Prize Foundation (\url{http://www.breakthroughinitiatives.org}). 

This material is based upon work supported in part by the National Science Foundation through Cooperative Agreements AST-1258333 and AST-2241526 and Cooperative Support Agreements AST-1202910 and 2211468 managed by the Association of Universities for Research in Astronomy (AURA), and the Department of Energy under Contract No. DE-AC02-76SF00515 with the SLAC National Accelerator Laboratory managed by Stanford University. Additional Rubin Observatory funding comes from private donations, grants to universities, and in-kind support from LSST-DA Institutional Members. This research uses services or data provided by the Rubin Science Platform at NSF-DOE Vera C. Rubin Observatory, which is jointly funded by the U.S. National Science Foundation and the U.S. Department of Energy, Office of Science. \update{This publication made use of proprietary Rubin Observatory Legacy Survey of Space and Time (LSST) data, and was prepared in accordance with the Rubin Observatory data rights and access policies. All authors of this publication meet the requirements for co-authorship of proprietary LSST data.}

 The authors acknowledge the efforts of the Rubin Construction Project, Commissioning, Data Management, Operations, and solar system Processing teams to build, commission, and bring online the Observatory and its software pipelines. We thank Matt Holman and Kevin Napier for their efforts to develop \texttt{pointdexter}. The authors thank Chris Lintott, Matt Holman, and Ken Smith for insightful discussions. We acknowledge the discussions on the Minor Planet Mailing List (MPML; \url{https://groups.io/g/mpml}) that helped inspire the brainstorming that eventually led to this work. \update{We thank the anonymous referee for their feedback that helped improve the final manuscript.} This research has made use of NASA’s Astrophysics Data System Bibliographic Services. The authors thank the Vera C. Rubin Observatory solar system Science Collaboration (SSSC) and especially the Layup Team and Sean O'Brien (who was supported by UK Science and Technology Facilities Council (STFC) grant ST/X001253/1) for helping to provide easy to use versions of the MPC reported Rubin photometry and astrometry and other calculated observational parameters.  
 
 This research has made use of the Science Explorer, funded by NASA under Cooperative Agreement 80NSSC21M00561. This research has made use of data and/or services provided by the International Astronomical Union's Minor Planet Center (MPC). Data from the MPC's database is made freely available to the public. Funding for this data and the MPC's operations comes from a NASA PDCO grant (80NSSC22M0024), administered via a University of Maryland - SAO subaward (106075-Z6415201). The MPC's computing equipment is funded in part by the above award, and in part by funding from the Tamkin Foundation.  We acknowledge the B612 Foundation's Asteroid Institute for providing accessible catalogs of the MPC records for Rubin Observatory small body submissions \citep{koumjian_2025_17047589}. This work made use of Astropy:\footnote{http://www.astropy.org} a community-developed core python package and an ecosystem of tools and resources for astronomy \citep{astropy2013,astropy2018,astropy2022}.  

Statement on the use of generative AI (Artificial Intelligence) tools: The authors used Claude AI for researching best practices for python programming, finding and summarizing python package capabilities and documentation, and to learn methods for optimizing/streamlining the analysis code. The manuscript text was written by the authors without the assistance or use of AI. 

Data Usage:  The main photometric and astrometry supporting this study are provided in tables accompanying this paper and are also available via the Minor Planet Center (\url{https://minorplanetcenter.net/}). Associated Rubin Observatory image data that the MPC reported photometry and astrometry were derived from will be released to  Rubin data rights holders as part of Rubin Observatory's Data Preview 2 \update{\citep[DP2; ][]{DP2,DP2techreport}} later in 2026. \update{DP2 catalogs and image data will be accessible in a public data release approximately in two years later after the full release of DP2 to Rubin data rights holders. }

\end{acknowledgments}



%
\facilities{Rubin:Simonyi (LSSTCam), Rubin:USDAC}

\software{Astropy \citep{astropy2013,astropy2018,astropy2022},Astroquery \citep{2019AJ....157...98G}, Color BLIndness Simulator (Colbis; https://www.color-blindness.com/coblis-color-blindness-simulator/), Claude Sonnet v4.6(Coding),  Jupyter Notebooks \citep{kluyver2016}, Layup \url{github.com/matthewholman/layup}, 
Matplotlib \citep{hunter2007}, Numpy \citep{harris2020}, Pandas \citep{mckinney2010, pandas2020}, Pooch \citep{uieda2020}, \texttt{REBOUND}\citep{rebound,reboundmercurius}, Small Body Dynamics Tool (SBDynT) \citep{Spencer:2026},  \texttt{sorcha} \citep{merritt2025}, sqlite3 (\url{https://docs.python.org/3/library/sqlite3.html}), tqdm \citep{dacostaluis2023}, Scipy \citep{2020SciPy-NMeth}
  }

\appendix 

\section{\update{Rubin Observations and Photometry Table}}

\update{\normalsize The MPC-reported Rubin observations used in this work are listed in Table \ref{tab:obs}. We use the photometry in the MPC instead of those available in Rubin Observatory's Data Preview 2 \citep[DP2; ][]{DP2, DP2techreport} to maximize the number of data points in our color analysis as not all observations have counterpart photometry and astrometry measurements in the DP2 source ($\ge$ 5-$\sigma$ detection in single frame LSSTCam observation) catalogs. The criteria for those images that are suitable to include in the Rubin data release and those that are suitable to submit detections to the MPC in the initial search for moving objects are different. The versions of the LSST Science Pipelines applied in each case differ slightly as well \citep{DP2techreport}. We note that for those observations that overlap the DP2 source catalog, we find consistent photometry within the reported measurement uncertainties. } 

\startlongtable
\begin{deluxetable}{c c c c c c c c c c c}
    \tabletypesize{\scriptsize}
    \tablecaption{Table of Rubin Observatory Observations \label{tab:obs}} 
    \tablewidth{\columnwidth}
    \tablehead{
        \colhead{Object} & \colhead{Observation} & \colhead{Obs. Midpoint} & \colhead{R.A} & \colhead{Dec} & \colhead{$m$} & \colhead{$\sigma_m$} & \colhead{Filter\tablenotemark{a}} & \colhead{$\alpha$} & \colhead{$r_H$} & \colhead{$\Delta$} \\
        \colhead{} & \colhead{Midpoint (UTC)} & \colhead{MJD (UTC)} & \colhead{(J2000)} & \colhead{(J2000)} & \colhead{(mag)} & \colhead{(mag)} & \colhead{} & \colhead{(deg)} & \colhead{(au)} & \colhead{(au)}
    }
    \startdata
    2025 MD138 & 2025 Jun 21.26 & 60847.25518 & 20:40:32.43 & $-$11:55:01.3 & 22.26 & 0.09 & $i$ & 1.3 & 29.582 & 28.796 \\
    2025 MD138 & 2025 Jun 21.30 & 60847.30434 & 20:40:32.21 & $-$11:55:01.8 & 22.19 & 0.08 & $i$ & 1.3 & 29.582 & 28.796 \\
    2025 MD138  & 2025 Jun 21.28 & 60847.27941 & 20:40:32.32 & $-$11:55:01.5 & 22.57 & 0.07 & $r$ & 1.3 & 29.582 & 28.796 \\
    2025 MD138  & 2025 Jul 18.36 & 60874.35746 & 20:37:57.60 & $-$12:02:19.3 & 22.18 & 0.08 & $i$ & 0.5 & 29.578 & 28.594 \\
    2025 MD138  & 2025 Jul 18.39 & 60874.38555 & 20:37:57.41 & $-$12:02:19.9 & 22.62 & 0.09 & $r$ & 0.5 & 29.578 & 28.594 \\
    2025 MD138  & 2025 Jul 18.10 & 60874.09820 & 20:37:59.28 & $-$12:02:13.8 & 22.99 & 0.07 & $g$ & 0.5 & 29.578 & 28.594 \\
    2025 MD138  & 2025 Jul 18.12 & 60874.12109 & 20:37:59.13 & $-$12:02:14.3 & 22.48 & 0.05 & $r$ & 0.5 & 29.578 & 28.594 \\
    2025 MD138  & 2025 Jul 18.12 & 60874.12160 & 20:37:59.13 & $-$12:02:14.3 & 22.60 & 0.06 & $r$ & 0.5 & 29.578 & 28.594 \\
    2025 MD138  & 2025 Jul 20.38 & 60876.37621 & 20:37:44.64 & $-$12:03:03.4 & 22.48 & 0.07 & $r$ & 0.5 & 29.578 & 28.594 \\
    2025 MD138  & 2025 Jul 20.35 & 60876.35270 & 20:37:44.80 & $-$12:03:02.9 & 22.88 & 0.07 & $g$ & 0.5 & 29.578 & 28.594 \\
    2025 MD138  & 2025 Jul 21.16 & 60877.15536 & 20:37:39.64 & $-$12:03:20.8 & 22.15 & 0.06 & $i$ & 0.5 & 29.578 & 28.594 \\
    2025 MD138  & 2025 Jul 21.16 & 60877.15731 & 20:37:39.63 & $-$12:03:20.8 & 22.10 & 0.05 & $i$ & 0.5 & 29.578 & 28.594 \\
    2025 MD138  & 2025 Jul 21.13 & 60877.13380 & 20:37:39.78 & $-$12:03:20.3 & 22.64 & 0.15 & $r$ & 0.5 & 29.578 & 28.594 \\
    2025 MD138  & 2025 Jul 29.36 & 60885.35600 & 20:36:46.11 & $-$12:06:34.0 & 22.18 & 0.06 & $i$ & 0.5 & 29.578 & 28.594 \\
    2025 MD138  & 2025 Jul 29.38 & 60885.37954 & 20:36:45.96 & $-$12:06:34.6 & 21.87 & 0.08 & $z$ & 0.5 & 29.578 & 28.594 \\
    \hline
    2025 MH348 & 2025 Jun 21.28 & 60847.27708 & 21:18:36.45 & $-$06:17:21.7 & 23.21 & 0.16 & $r$ & 1.5 & 29.805 & 29.153 \\
    2025 MH348 & 2025 Jul 01.31 & 60857.31340 & 21:17:54.71 & $-$06:20:11.8 & 23.38 & 0.11 & $r$ & 1.5 & 29.805 & 29.153 \\
    2025 MH348 & 2025 Jul 02.22 & 60858.22247 & 21:17:50.50 & $-$06:20:30.3 & 23.14 & 0.10 & $r$ & 1.5 & 29.805 & 29.153 \\
    2025 MH348 & 2025 Jul 18.24 & 60874.23615 & 21:16:26.49 & $-$06:27:22.2 & 23.39 & 0.12 & $r$ & 1.3 & 29.806 & 29.035 \\
    2025 MH348 & 2025 Jul 18.26 & 60874.25674 & 21:16:26.38 & $-$06:27:22.7 & 22.94 & 0.13 & $i$ & 1.3 & 29.806 & 29.035 \\
    2025 MH348 &  2025 Jul 20.21 & 60876.21262 & 21:16:15.07 & $-$06:28:23.2 & 23.88 & 0.12 & $g$ & 1.3 & 29.806 & 29.035 \\
    2025 MH348 & 2025 Jul 20.21 & 60876.21357 & 21:16:15.06 & $-$06:28:23.2 & 23.77 & 0.10 & $g$ & 1.3 & 29.806 & 29.035 \\
    2025 MH348 &  2025 Jul 21.15 & 60877.15263 & 21:16:09.59 & $-$06:28:52.9 & 22.69 & 0.14 & $i$ & 1.3 & 29.806 & 29.035 \\
    2025 MH348 & 2025 Jul 22.26 & 60878.26157 & 21:16:03.03 & $-$06:29:28.6 & 22.84 & 0.07 & $i$ & 1.3 & 29.806 & 29.035 \\
    2025 MH348 & 2025 Jul 23.14 & 60879.14297 & 21:15:57.80 & $-$06:29:57.4 & 23.25 & 0.11 & $r$ & 1.3 & 29.806 & 29.035 \\
    2025 MH348 & 2025 Jul 24.24 & 60880.23976 & 21:15:51.23 & $-$06:30:33.8 & 23.54 & 0.10 & $g$ & 1.3 & 29.806 & 29.035 \\
    \hline
    2025 NN80  & 2025 Jul 07.11 & 60863.10998 & 21:04:55.46 & $-$20:07:48.1 & 22.09 & 0.17 & $i$ & 1.4 & 28.333 & 27.586 \\
    2025 NN80  & 2025 Jul 07.11 & 60863.10772 & 21:04:55.49 & $-$20:07:47.7 & 22.04 & 0.17 & $i$ & 1.4 & 28.333 & 27.586 \\
    2025 NN80  & 2025 Jul 07.13 & 60863.13406 & 21:04:55.33 & $-$20:07:49.3 & 22.16 & 0.19 & $z$ & 1.4 & 28.333 & 27.586 \\
    2025 NN80  &  2025 Jul 07.14 & 60863.14067 & 21:04:55.31 & $-$20:07:49.5 & 21.74 & 0.14 & $z$ & 1.4 & 28.333 & 27.586 \\
    2025 NN80  & 2025 Jul 18.38 & 60874.38321 & 21:03:45.30 & $-$20:15:41.7 & 22.24 & 0.06 & $r$ & 1.1 & 28.335 & 27.474 \\
    2025 NN80  & 2025 Jul 18.36 & 60874.35500 & 21:03:45.48 & $-$20:15:40.6 & 22.14 & 0.07 & $i$ & 1.1 & 28.335 & 27.474 \\
    2025 NN80  &  2025 Jul 20.35 & 60876.34820 & 21:03:32.35 & $-$20:17:05.4 & 22.69 & 0.07 & $g$ & 1.1 & 28.335 & 27.474 \\
    2025 NN80  & 2025 Jul 20.35 & 60876.34864 & 21:03:32.34 & $-$20:17:05.5 & 22.94 & 0.08 & $g$ & 1.1 & 28.335 & 27.474 \\
    2025 NN80  &  2025 Jul 20.37 & 60876.37212 & 21:03:32.18 & $-$20:17:06.6 & 22.35 & 0.05 & $r$ & 1.1 & 28.335 & 27.474 \\
    2025 NN80  &  2025 Jul 20.30 & 60876.29683 & 21:03:32.69 & $-$20:17:03.3 & 22.80 & 0.08 & $g$ & 1.1 & 28.335 & 27.474 \\
    2025 NN80  &  2025 Jul 20.37 & 60876.37168 & 21:03:32.19 & $-$20:17:06.5 & 22.34 & 0.05 & $r$ & 1.1 & 28.335 & 27.474 \\
    2025 NN80  &  2025 Jul 22.37 & 60878.37334 & 21:03:18.81 & $-$20:18:32.0 & 22.18 & 0.10 & $r$ & 1.1 & 28.335 & 27.474 \\
    2025 NN80  & 2025 Jul 22.39 & 60878.39438 & 21:03:18.68 & $-$20:18:33.1 & 22.93 & 0.14 & $g$ & 1.1 & 28.335 & 27.474 \\
    2025 NN80  &  2025 Jul 29.35 & 60885.34587 & 21:02:31.13 & $-$20:23:30.1 & 21.98 & 0.04 & $i$ & 1.0 & 28.335 & 27.434 \\
    2025 NN80  & 2025 Jul 29.37 & 60885.36951 & 21:02:30.97 & $-$20:23:31.1 & 21.96 & 0.07 & $z$ & 1.0 & 28.335 & 27.434 \\
    2025 NN80  & 2025 Jul 29.38 & 60885.37541 & 21:02:30.93 & $-$20:23:31.3 & 21.98 & 0.07 & $z$ & 1.0 & 28.335 & 27.434 \\
    2025 NN80  & 2025 Jul 29.35 & 60885.35188 & 21:02:31.09 & $-$20:23:30.3 & 22.13 & 0.04 & $i$ & 1.0 & 28.335 & 27.434 \\
    2025 NN80  & 2025 Jun 21.24 & 60847.23998 & 21:06:17.59 & $-$19:57:27.2 & 22.30 & 0.07 & $i$ & 1.4 & 28.333 & 27.586 \\
    2025 NN80  & 2025 Jun 21.26 & 60847.26384 & 21:06:17.49 & $-$19:57:28.0 & 22.46 & 0.05 & $r$ & 1.4 & 28.333 & 27.586 \\
    2025 NN80  & 2025 Jun 21.26 & 60847.26335 & 21:06:17.49 & $-$19:57:28.0 & 22.65 & 0.06 & $r$ & 1.4 & 28.333 & 27.586 \\
    2025 NN80  & 2025 Jul 02.26 & 60858.25533 & 21:05:22.97 & $-$20:04:31.1 & 22.44 & 0.06 & $r$ & 1.4 & 28.333 & 27.586 \\
    2025 NN80  & 2025 Jul 02.24 & 60858.24144 & 21:05:23.05 & $-$20:04:30.5 & 22.18 & 0.10 & $i$ & 1.4 & 28.333 & 27.586 \\
    2025 NN80  & 2025 Jul 02.24 & 60858.24195 & 21:05:23.05 & $-$20:04:30.5 & 22.18 & 0.07 & $i$ & 1.4 & 28.333 & 27.586 \\
    2025 NN80  & 2025 Jul 04.19 & 60860.19332 & 21:05:12.24 & $-$20:05:49.3 & 22.11 & 0.11 & $z$ & 1.4 & 28.333 & 27.586 \\
    2025 NN80  & 2025 Jul 22.11 & 60878.11315 & 21:03:20.58 & $-$20:18:21.0 & 22.02 & 0.07 & $z$ & 1.1 & 28.335 & 27.474 \\
    2025 NN80  & 2025 Aug 11.31 & 60898.30978 & 21:01:00.69 & $-$20:32:31.8 & 22.05 & 0.11 & $i$ & 0.6 & 28.337 & 27.365 \\
    2025 NN80  & 2025 Aug 11.31 & 60898.30933 & 21:01:00.69 & $-$20:32:31.7 & 21.82 & 0.09 & $i$ & 0.6 & 28.337 & 27.365 \\
    2025 NN80  & 2025 Aug 11.35 & 60898.34803 & 21:01:00.42 & $-$20:32:33.4 & 21.74 & 0.14 & $z$ & 0.6 & 28.337 & 27.365 \\
    2025 NN80  & 2025 Aug 11.32 & 60898.32157 & 21:01:00.60 & $-$20:32:31.8 & 22.00 & 0.19 & $z$ & 0.6 & 28.337 & 27.365 \\
    2025 NN80  & 2025 Aug 17.24 & 60904.23846 & 21:00:19.81 & $-$20:36:28.7 & 22.89 & 0.07 & $g$ & 0.3 & 28.339 & 27.331 \\
    2025 NN80  & 2025 Aug 17.29 & 60904.28505 & 21:00:19.48 & $-$20:36:30.7 & 22.25 & 0.11 & $i$ & 0.3 & 28.339 & 27.331 \\
    2025 NN80  & 2025 Aug 17.32 & 60904.31546 & 21:00:19.24 & $-$20:36:32.3 & 21.84 & 0.12 & $i$ & 0.3 & 28.339 & 27.331 \\
    \enddata
    \tablenotetext{a}{LSST photometric system}
     \tablenotetext{}{Phase angle = $\alpha$; heliocentric distance = $r_H$; geocentric distance = $\Delta$}
     \tablenotetext{}{All observations had 30s exposure times.}
    \tablenotetext{}{\update{Photometry was retrieved from the MPC on 2026 12 June}}
      \tablenotetext{}{Phase angle, heliocentric distance, and geocentric distance values are from queries to the JPL Horizons \citep{giorgini1996} on 2026 12 June}
     \tablenotetext{}{ 2025 MH348 and 2025 NN80 are confirmed as dynamically stable Neptune L5 Trojans. 2025 MD138 is a candidate stable Neptune L5 Trojan (See Section \ref{sec:orbits}).}
     \tablenotetext{}{This table is available in machine-readable form in the online article}
\end{deluxetable}

\section{\update{Photometric transformations}}
\label{app:photometric}

\update{Following \cite{RTN-099}, we have that the transformations between the DES photometric system and the LSST system are:}
\begin{align}
    g_{\rm LSST} & =  g_{\rm DES} + 0.016 (g-i)_{\rm DES} - 0.003 (g-i)^2_{\rm DES} + 0.006, \\
    r_{\rm LSST} & = r_{\rm DES} + 0.185 (r-i)_{\rm DES} - 0.015 (r-i)^2_{\rm DES} + 0.010, \\
    i_{\rm LSST} & = i_{\rm DES} + 0.150 (r-i)_{\rm DES} - 0.003 (r-i)^2_{\rm DES} - 0.009,\\
    z_{\rm LSST} & =  z_{\rm DES} + 0.270 (i-z)_{\rm DES} + 0.036 (i-z)^2_{\rm DES} - 0.003.
\end{align}
\update{Note that these transformations have a scatter of $\lesssim 10\,\mathrm{mmag}$. Since none of our objects (either in the Rubin sample or in the DES set) have measurements on these level of precision, we are safe to ignore this additional source of error.}

\update{To simplify our transformations, we use these to derive LSST colors as a function of the DES ones:}
\begin{align}
    (g-r)_{\rm LSST}  = & (g-r)_{\rm DES} + 0.016 (g-i)_{\rm DES} - 0.185 (r-i)_{DES}  - 0.003 (g-i)^2_{\rm DES} \\\nonumber  & + 0.015 (r-i)_{\rm DES}^2 - 0.004, \\ 
    (r-i)_{\rm LSST} = & 1.035 (r-i)_{\rm DES} - 0.012 (r-i)^2_{\rm DES} + 0.019  \\
     (r-z)_{\rm LSST}  = & (r-z)_{\rm DES} + 0.185 (r-i)_{\rm DES} - 0.270 (i-z)_{\rm DES} \\ \nonumber & - 0.015 (r-i)^2_{\rm DES} - 0.036 (i-z)_{\rm DES}^2 + 0.013.
\end{align} 

\update{Similarly, the transformations between the LSST and SDSS systems are}
\begin{align} 
    g_{\rm LSST} & = g_{\rm SDSS} - 0.062 (g-i)_{\rm SDSS} + 0.007, \\  
    r_{\rm LSST} & = r_{\rm SDSS} - 0.007 (g-i)_{\rm SDSS} + 0.005, \\
    i_{\rm LSST} & = i_{\rm SDSS} - 0.010 (g-i)_{\rm SDSS} + 0.019, \\
    z_{\rm LSST} & = z_{\rm SDSS} + 0.020 (g-i)_{\rm SDSS} + 0.032.
\end{align}

\update{As before, we can write these as color transformations:}
\begin{align}
    (g-r)_{\rm LSST} & =  (g-r)_{\rm SDSS} - 0.069 (g-i)_{\rm SDSS} + 0.002, \\
    (r-i)_{\rm LSST} & =  (r-i)_{\rm SDSS} + 0.003(g-i)_{\rm SDSS} - 0.014, \\
    (r-z)_{\rm LSST} & =  (r-z)_{\rm SDSS} + 0.013(g-i)_{\rm SDSS} - 0.027.
\end{align}

\update{We note that at the time of publication, these represent the most up to date color transforms available for comparing observations in non-LSST filters to LSSTCam photometry \citep{DP2}. We caution the reader that \cite{RTN-099} describes their work as a living document; as their filter transformations were estimated based off of  ComCam commissioning data, Rubin's Data Preview 1 \citep[DP1;][]{DP1,2026AJ....171..360V}, these will likely be refined as the LSST progresses. }


\bibliography{ms}{}
\bibliographystyle{aasjournalv7}



\end{document}